\documentclass[journal,onecolumn,10pt,draftclsnofoot]{IEEEtran}
\usepackage{comment}
\usepackage[utf8]{inputenc}
\usepackage{tabularx}
\usepackage{hyperref}
\usepackage{stackrel}
\usepackage{amssymb}
\usepackage{amsmath}
\usepackage{mathtools}
\usepackage[dvips]{graphics}
\usepackage{graphicx}
\usepackage{psfrag}
\usepackage{epsf}
\usepackage{dsfont}
\usepackage{stackrel}
\usepackage{theorem}
\usepackage{bigints}
\usepackage{arydshln}
\usepackage{fancyvrb}
\usepackage{geometry}
\usepackage{pstool}
\usepackage{subcaption}
\usepackage{tabularx}
\usepackage{subcaption}
\usepackage[noadjust]{cite} 			
\newtheorem{definition}{Definition}
\newtheorem{theorem}{Theorem}

\newtheorem{lemma}{Lemma}

\newcommand{\defeq}{ \triangleq }

\newcommand{\Typical}{\mathcal{A}^{(n)}}

\renewcommand{\Pr}{\mathbb{P}}
\newcommand{\indicate}{\mathds{1}}
\newcommand{\expectation}{\mathbb{E}}
\newcommand{\Markov}{\leftrightarrow}
\newcommand{\dprime}{{\prime\prime}}
\newcommand\independent{\protect\mathpalette{\protect\independenT}{\perp}}
\def\independenT#1#2{\mathrel{\rlap{$#1#2$}\mkern2mu{#1#2}}}

\begin{document}
	\title{Cooperative Binning for Semi-deterministic Channels with Non-causal State Information}
	\author{~Ido~B.~Gattegno,~\IEEEmembership{Student~Member,~IEEE,} 
	~Haim~H.~Permuter,~\IEEEmembership{Senior~Member,~IEEE,} 
	~Shlomo ~Shamai ~(Shitz),~\IEEEmembership{Fellow,~IEEE}
	and~Ayfer~\"{O}zg\"{u}r,~\IEEEmembership{Member,~IEEE}
	\thanks{The work of Ido B. Gattegno, Haim H. Permuter and Shlomo Shamai was supported by the Heron consortium via the minister of economy and science, and, by the ERC (European Research Council). The work of A. Ozgur was supported in part by NSF grant \#1514538.}}
	\maketitle
	\begin{abstract}\label{Section: Abstract}
		The capacity of the semi-deterministic relay channel (SD-RC) with \emph{non-causal} channel state information (CSI) only at the encoder and decoder is characterized. 
		The capacity is achieved by a scheme based on \emph{cooperative-bin-forward}. 
		This scheme allows cooperation between the transmitter and the relay without the need to decode a part of the message by the relay. 
		The transmission is divided into blocks and each deterministic output of the channel (observed by the relay) is mapped to a bin. 
		The bin index is used by the encoder and the relay to choose the cooperation codeword in the next transmission block.
		In \emph{causal} settings the cooperation is independent of the state. In \emph{non-causal} settings dependency between the relay's transmission and the state can increase the transmission rates.
		The encoder implicitly conveys partial state information to the relay. In particular, it uses the states of the next block and selects a cooperation codeword accordingly
		and the relay transmission depends on the cooperation codeword and therefore also on the states. 
		We also consider the multiple access channel with partial cribbing as a semi-deterministic channel.
		The capacity region of this channel with non-causal CSI is achieved by the new scheme. 
		Examining the result in several cases, we introduce a new problem of a point-to-point (PTP) channel where the state is provided to the transmitter by a state encoder. 
		Interestingly, even though the CSI is also available at the receiver, we provide an example which shows that the capacity with non-causal CSI at the state encoder is strictly larger than the capacity with causal CSI.
	\end{abstract}
	\section{Introduction}\label{Section: introduction}
	\par Semi-deterministic models describe a variety of communication problems in which there exists a deterministic link between a transmitter and a receiver. 
	This work focus on the semi-deterministic relay channel (SD-RC) and the multiple access channel (MAC) with partial cribbing encoders and non-causal channel state information (CSI) only at the encoder and decoder.
	The state of a channel may be governed by physical phenomena or by an interfering transmission over the channel, and the deterministic link may also be a function of this state. 
	\par The capacity of the relay channel was first studied by van der Muelen \cite{meulen1971relaychannels}. In the relay channel, an encoder receives a message, denoted by $M$, and sends it to a decoder over a channel with two outputs. A relay observes one of the channel outputs, denoted by $Z$, and uses past observations in order to help the encoder deliver the message. The decoder observes the other output, denoted by $Y$, and uses it to decode the message that was sent by the encoder.
	Cover and El-Gamal \cite{coverel1979capacityrelay} established achievable rates for the general relay channel, using a \emph{partial-decode-forward} scheme. 
	If the channel is semi-deterministic (i.e. the output to the relay is a function of the channel inputs), El-Gamal and Aref \cite{el1982sdrc} showed that this scheme achieves the capacity. Partial-decode-forward operates as follows: first, the transmission is divided into $B$ blocks, each of length $n$; in each block $b$ we send a message $M^{(b)}$, at rate $R$, that is independent of the messages in the other blocks. 
	The message is split; after each transmission block, the relay decodes a part of the message and forwards it to the decoder in the next block using its transmission sequence. Since the encoder also knows the message, it can cooperate with the relay in the next block. The capacity of the SD-RC is given by maximizing $\min \left\{I(X,X_r;Y),H(Z|X_r) + I(X;Y|X_r,Z)\right\}$ over the joint probability mass function (PMF) $p_{X,X_r}$, where $X$  is the input from the encoder and $X_r$ is the input from the relay. The cooperation is expressed in the joint PMF, in which $X$ and $X_r$ are dependent. However, when the channel depends on a state that is unknown to the relay, the partial-decode-forward scheme is suboptimal \cite{kolte2016binning}, i.e., it does not achieve the capacity. The partial-decoding procedure at the relay is too restrictive since the relay is not aware of the channel state.
	\begin{figure}[t]
		\centering
		\psfragscanon
		\psfragfig*[mode=nonstop,scale=0.5]{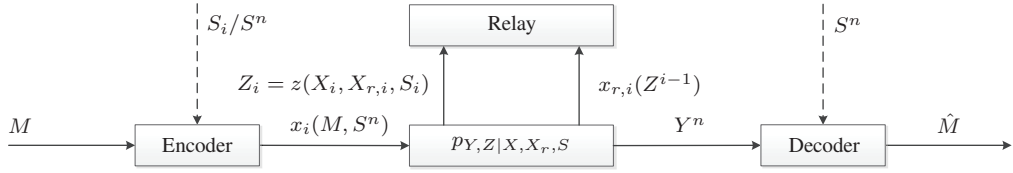}{
			\psfrag{A}[][][1]{\footnotesize $M$}
			\psfrag{B}[][][1]{\footnotesize \hspace*{3mm}$S_i/S^n$}
			\psfrag{C}[][][1]{\footnotesize Encoder}
			\psfrag{D}[][][1]{\footnotesize $x_{i}(M,S^n)$}
			\psfrag{E}[][][1]{\footnotesize $\hspace{-20mm}Z_i=z(X_i,X_{r,i},S_i)$}
			\psfrag{F}[][][1]{\footnotesize Relay}
			\psfrag{G}[][][1]{\footnotesize $\hspace{10mm}x_{r,i}(Z^{i-1})$}
			\psfrag{H}[][][1]{\footnotesize $p_{Y,Z|X,X_{r},S}$}
			\psfrag{I}[][][1]{\footnotesize $Y^n$}
			\psfrag{J}[][][1]{\footnotesize $S^n$}
			\psfrag{K}[][][1]{\footnotesize Decoder}
			\psfrag{L}[][][1]{\footnotesize \hspace*{5mm} $\hat{M}$}
		}
		\psfragscanoff
		\caption{SD-RC with causal/non-causal CSI at encoder and decoder.}
		\label{Figure: SD-RC with causal and non-causal states}
	\end{figure}
	\par Focusing on state-dependent SD-RC, depicted in Fig. \ref{Figure: SD-RC with causal and non-causal states}, we consider two situations: when the CSI is available in a \emph{causal} or a \emph{non-causal} manner. 
	State-dependent relay channels were studied in \cite{akhbari2010compressforward,muji2013relaystates,muji2008relaycausalstate,deng2013orthogonalrelay,zaidi2013relaystatesource,zaifdi2010cooperativerelaying,kim2008classdet,aguerri2012orthogonalrelay,kolte2016binning}; 
	Kolte et al. \cite{kolte2016binning} derived the capacity of state-dependent SD-RC with \emph{causal} CSI and introduced a \emph{cooperative-bin-forward} coding scheme. 
	It differs from {partial-decode-forward} as follows: the relay does not have to explicitly recover the message bits; instead, the encoder and relay agree on a map from the deterministic outputs space $\mathcal{Z}^n$ to a bin index. 
	This index is used by the relay to choose the next transmission sequence. Note that this \emph{cooperative-binning} is independent of the state and, therefore, can be used by the relay. 
	The encoder is also aware of this index (since the output is deterministic) and coordinates with the relay in the next block, despite the lack of state information at the relay. The capacity of this channel is given by maximizing $\min \left\{I(X,X_r;Y|S),H(Z|X_r,S) + I(X;Y|X_r,Z,S)\right\}$ over $p_{X_r}p_{X|X_rS}$. Note that $X$ and $X_r$ are dependent, but $X_r$ and $S$ are not. When the state is known causally, a dependency between $X_r$ and $S$ is not feasible. At each time $i$, the encoder can send to the relay information about the states up to time $i$. The relay can use only \emph{strictly causal} observations $Z^{i-1}$, which may contain information on $S^{i-1}$ but not on $S_i$. Furthermore, since the states are distributed independently, the past state at the relay does not help to increase the achievable rate.
	\par The main contribution of this paper is to develop a variation of the cooperative-bin-forward scheme that accounts for non-causal CSI.
	While the former scheme allows cooperation, the new scheme also allows dependency between the relay's transmission and the state.
	When the CSI is available in a \emph{non-causal} manner, knowledge of the state at the relay is feasible and may increase the transmission rate. 
	The encoder can perform a look-ahead operation and transmit to the relay information about the upcoming states. The relay can still agree with the encoder on a map, and in each transmission the encoder can choose carefully which index it causes the relay to see. The encoder chooses an index such that it reveals compressed state information to the relay, using an \emph{auxiliary} cooperation codeword. Incorporating look-ahead operations with cooperative-binning increases the transmission rate and achieves capacity. This scheme can be used in other semi-deterministic models, such as the multiple access channel (MAC) with strictly causal partial cribbing and non-causal CSI. 
	\par The MAC with cooperation can also be viewed as a semi-deterministic model, due the deterministic cooperation link. MAC with conferencing, introduced by Willems in \cite{willems83mac_cooperation},  consists of a rate-limited private link between two encoders. 
	Permuter \emph{el al} \cite{permuter2011confwstate} showed that for \emph{state-dependent} MAC with conferencing, the capacity can be achieved by superposition coding and rate-splitting. 
	The cribbing is a different type of cooperation, also introduced by Willems \cite{willems85cribbing}, in which one transmitter has access to (is cribbing) the transmission of the other. In \cite{simeone2012cooperative}, Simeone \emph{et al.} considered cooperative wireless cellular systems and analyzed their performance with cribbing (referred to as In-Band cooperation). 
	The results show how cribbing potentially increases the capacity. 
	A generalization of the cribbing is partial and controlled cribbing, introduced by Asnani and Permuter in \cite{Asnani2011contrcrib}, when one encoder has \emph{limited} access to the transmission sequence of the other. 
	The cribbed information is a deterministic function of the transmission sequence.
	Kopetz \emph{et al.} \cite{kopetz2016partialcribbing} characterized the capacity region of combined \emph{partial} cribbing and conferencing MAC without states. 
	When states are known \emph{causally} at the first encoder (while the other is cribbing),  Kolte \emph{et al.} \cite{kolte2016binning} derived the capacity, which is achieved by cooperative-bin-forward. 
	We show that the variation of the cooperative-bin-forward scheme achieves the capacity when the states are known \emph{non-causally}.
	\par The results are examined for several special cases; the first is a point-to-point (PTP) channel where the CSI is available to the transmitter through a state encoder, and to the receiver. 
	Former work on limited CSI was done by Rosenzweig \emph{el al} \cite{rosenweig2005partialcsi}, where the link from the state encoder to the transmitter is rate-limited. 
	Steinberg \cite{steinberg2008ratelimited} derived the capacity of rate-limited state information at the receiver.
	In our setting, the link between the state encoder and the transmitter is not is not a rate-limited bit pipe, but a communication channel where the transmitter can observe the output of the state encoder in a causal fashion. 
	We provide an example which illustrates that in this setting the capacity with non-causal CSI available at the state encoder is strictly larger than the capacity with causal CSI at the state encoder, even-though the receiver also has channel state information. This is somewhat surprising given that in a PTP channel the CSI at both the transmitter and receiver, causal and non-causal state information lead to the same capacity.
	\par The remainder of the paper is organized follows.
	Problem definitions and capacity theorems are given in Section \ref{Section: Main Results}.
	Special cases are given in Section \ref{Section: Special cases}, and the new state-encoder problem and the example are given in Section \ref{Section: PTP w SE}.
	Proofs for theorems are given in Sections \ref{Section: SD-RC capacity},\ref{Section: Proof MAC with one state} and \ref{Section: Proof for MAC with two states}. 
	In Section \ref{Section: Conclusion} we offer conclusions and final remarks.
	\section{Problem Definition and Main Results}\label{Section: Main Results}
	\subsection{Notation}
	\par We use the following notation. Calligraphic letters denote discrete sets, e.g., $\mathcal{X}$. 
	Lowercase letters, e.g., $x$, represent variables. A vector of $n$ variables $(x_1,\dots,x_n)$ is denoted by $x^n$. A substring of $x^n$ is denoted by $x_i^j$, and includes variables $(x_i,\dots,x_j)$.
	Whenever the dimensions are clear from the context, the subscript is omitted. Let $(\Omega,\mathcal{F},\Pr)$ denote a probability space where $\Omega$ is the sample space, $\mathcal{F}$ is the $\sigma$-algebra and $\Pr$ is the probability measure. Roman face letters denote events in the $\sigma$-algebra, e.g., $\mathrm{A}\in\mathcal{F}$.  $\Pr\left[\mathrm{A}\right]$ is the probability assigned to $\mathrm{A}$, and $\indicate[\mathrm{A}]$ is the indicator function, i.e., indicates if event $\mathrm{A}$ has occurred. Random variables are denoted by uppercase letters, e.g., $X$, and similar conventions apply for vectors.
	The probability mass function (PMF) of a random variable, $X$, is denoted by $p_X$. If $x\in\mathcal{X}$, then $p_X(x)=\Pr\left[X=x\right]$. Whenever the random variable is clear from the context, we drop the subscript. Similarly, a joint distribution of $X$ and $Y$ is denoted by $p_{X,Y}$ and a conditional PMF by $p_{Y|X}$. Whenever $Y$ is a deterministic function of $X$, we denote $Y=f(X)$ and the conditional PMF by $1_{Y|X}$. If $X$ and $Y$ are independent, we denote this as $X\independent Y$ which implies that $p_{X,Y}=p_Xp_Y$,  and a Markov chain is denoted as $X\Markov Y \Markov Z$ and implies that $p_{X,Y,Z}=p_{X,Y}p_{Z|Y}$.
	\par An empirical mass function (EMF) is denoted by $\nu(a|x^n)=\frac{1}{n}\sum_{i=1}^{n}\indicate[x_i=a]$.
	Sets of typical sequences are denoted by $\Typical_\epsilon(p_X)$, which is a $\epsilon$-strongly typical set with respect to PMF $p_X$, and defined by
	\begin{align}
		\Typical_\epsilon(p_X)\defeq\left\{x^n : \; |\nu(a|x^n) - p_X(a) | < \epsilon p_X(a) ,\; \forall a \in \mathcal{X}  \right\}.
	\end{align}
	Jointly typical sets satisfy the same definition with respect to (w.r.t.) the joint distribution and are denoted by  $\Typical_\epsilon(p_{X,Y})$. Conditional typical sets are defined as
	\begin{align}
		\Typical_\epsilon(p_{X,Y}\vert y^n)\defeq\left\{x^n : \; (x^n,y^n)\in\Typical_\epsilon(p_{X,Y})  \right\}.
	\end{align}
	\subsection{Semi-Deterministic Relay Channel}\label{Subsection: Main Results - SD-RC definition}
		\begin{figure}[t]
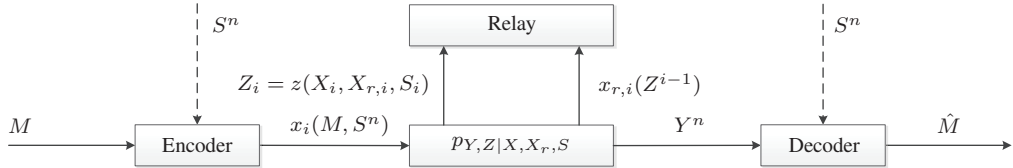

			\centering
			\psfragscanon
			\psfragfig*[mode=nonstop,scale=0.5]{SD_RC}{
				\psfrag{A}[][][1]{\footnotesize $M$}
				\psfrag{B}[][][1]{\footnotesize $S^n$}
				\psfrag{C}[][][1]{\footnotesize Encoder}
				\psfrag{D}[][][1]{\footnotesize $x_{i}(M,S^n)$}
				\psfrag{E}[][][1]{\footnotesize $\hspace{-20mm}Z_i=z(X_i,X_{r,i},S_i)$}
				\psfrag{F}[][][1]{\footnotesize Relay}
				\psfrag{G}[][][1]{\footnotesize $\hspace{10mm}x_{r,i}(Z^{i-1})$}
				\psfrag{H}[][][1]{\footnotesize $p_{Y,Z|X,X_{r},S}$}
				\psfrag{I}[][][1]{\footnotesize $Y^n$}
				\psfrag{J}[][][1]{\footnotesize $S^n$}
				\psfrag{K}[][][1]{\footnotesize Decoder}
				\psfrag{L}[][][1]{\footnotesize \hspace*{5mm} $\hat{M}$}
			}
			\psfragscanoff
			\caption{SD-RC with non-causal CSI at encoder and decoder.}.
			\label{Figure: SD-RC with non-causal states}
		\end{figure}
		\par We begin with a state dependent SD-RC, depicted in Fig. \ref{Figure: SD-RC with non-causal states}. This channel depends on a state $S_i\in\mathcal{S}$, which is known non-causally to the encoder and decoder, but not to the relay. An encoder sends a message $M$ to the decoder through a channel with two outputs. The relay observes an output $Z^n$ of the channel, which at time $i$ is a deterministic function of the channel inputs, $X_i$ and $X_{r,i}$, and the state (i.e., $Z_i=z(X_i,X_{r,i},S_i)$). Based on past observations $Z^{i-1}$ the relay transmits $X_{r,i}$ in order to assist the encoder. The decoder uses the state information and the channel output $Y^n$ in order to estimate $\hat{M}$. The channel is memoryless and characterized by the joint PMF $p_{Y,Z|X,X_r,S} = 1_{Z|X,X_r,S}p_{Y|Z,X,X_r,S}$.
		\begin{definition}[Code for SD-RC]
			\par A $(R,n)$ code $\mathcal{C}_n$ for the SD-RC is defined by
			\begin{align*}
				x^n:	&	\left[1:2^{nR}\right]\times \mathcal{S}^n \to \mathcal{X}^n\\
				x_{r,i}:&	\mathcal{Z}^{i-1} \to \mathcal{X}_r	& 1\leq i \leq n \\
				\hat{m}:& 	\mathcal{Y}^n\times\mathcal{S}^n \to \left[1:2^{nR}\right]&
			\end{align*}
		\end{definition}   
		\begin{definition}[Achievable rate]
			A rate $R$ is said to be achievable if there exists $(R,n)$ such that
			\begin{align}
				P_e(\mathcal{C}_n)\defeq\Pr_{\mathcal{C}_n}\left[\hat{m}(Y^n,S^n)\neq M\right] \leq \epsilon
			\end{align}
			for any $\epsilon>0$ and some sufficiently large $n$. 
		\end{definition}
		The capacity is defined to be the supremum of all achievable rates.
		\begin{theorem}\label{Theorem: SD-RC non-causal capacity}
			The capacity of the SD-RC with non-causal CSI, depicted in Figure \ref{Figure: SD-RC with non-causal states}, is given by
			\begin{align}\label{Equation: SD-RC capacity non-causal}
			C=\max \min\left\{I(X,X_r;Y|S),I(X;Y|X_r,Z,S,U) + H(Z|X_r,S,U) - I(U;S)\right\}
			\end{align}
			where the maximum is over $p_{U|S}p_{X_r|U}p_{X|X_r,U,S}$ such that $I(U;S)\leq H(Z|X_r,S,U)$, where $Z=z(X,X_r,S)$ and $|\mathcal{U}| \leq \min\{|\mathcal{S}||\mathcal{X}||\mathcal{X}_r|,|\mathcal{S}||\mathcal{Y}|+1\}$.
		\end{theorem}
		\par The proof for the theorem is given in Section \ref{Section: SD-RC capacity}. 
		Let us first investigate the capacity and the role of the auxiliary random variable $U$.
		Here, the random variable $U$ is used to create empirical coordination between the encoder, the relay and the states, i.e., with high probability $(S^n,U^n,X_r^n,X^n)$ are jointly typical w.r.t. $p_{S,U,X_r,X}$. 
		Note that the PMF factorizes as $p_{U|S}p_{X_r|U}p_{X|X_r,U,S}$; the random variable $X_r$, which represents the relay, depends on $S$ through the random variable $U$. 
		This dependency represents the state knowledge at the relay, using an auxiliary codeword $U^n$. 
		\subsection{Multiple Access Channel with Partial Cribbing}\label{Subsection: Main Results - MAC with partial cribbing}
		\par Consider a MAC with partial cribbing and non-causal state information, as depicted in Figure \ref{Figure: DM-MAC with one side cribbing and non-causal states, two states}. This channel depends on the state $(S_1,S_2)$ sequence that is known to the decoder, and each encoder $w\in\{1,2\}$ has non-causal access to one state component $S_{w}\in\mathcal{S}_w$. Each encoder $w$ sends a message $M_w$ over the channel. Encoder 2 is cribbing Encoder 1; the cribbing is strictly causal, partial and controlled by $S_1$. Namely, the cribbed signal at time $i$, denoted by $Z_i$, is a deterministic function of $X_{1,i}$ and $S_{1,i}$. The cribbed information is used by Encoder 2 to assist Encoder 1.
		\begin{figure}[t]
			\centering
			\psfragscanon
			\psfragfig*[mode=nonstop,scale=0.5]{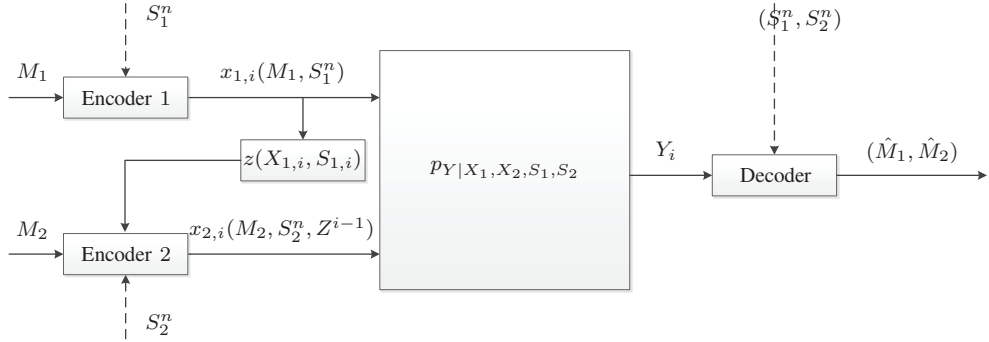}{
				\psfrag{A}[][][1]{\footnotesize $M_1$}
				\psfrag{B}[][][1]{\footnotesize $M_2$}
				\psfrag{C}[][][1]{\footnotesize $S_1^n$}
				\psfrag{D}[][][1]{\footnotesize Encoder $1$}
				\psfrag{E}[][][1]{\footnotesize Encoder $2$}
				\psfrag{F}[][][1]{\footnotesize $x_{1,i}(M_1,S_1^n)$}
				\psfrag{G}[][][1]{\footnotesize $z(X_{1,i},S_{1,i})$}
				\psfrag{H}[][][1]{\footnotesize $x_{2,i}(M_2,S_2^n,Z^{i-1})$}
				\psfrag{I}[][][1]{\footnotesize $p_{Y|X_1,X_2,S_1,S_2}$}
				\psfrag{J}[][][1]{\footnotesize $Y_i$}
				\psfrag{K}[][][1]{\footnotesize $(S_1^n,S_2^n)$}			
				\psfrag{L}[][][1]{\footnotesize Decoder}
				\psfrag{M}[][][1]{\footnotesize $(\hat{M}_1,\hat{M}_2)$}	
				\psfrag{O}[][][1]{\footnotesize $S_2^n$}
			}
			\psfragscanoff
			\caption{State dependent MAC with two state components and one side cribbing. The cribbing is \emph{strictly} causal -- $X_2=x_{2,i}(M_2,S_2^n,Z^{i-1})$.}
			\label{Figure: DM-MAC with one side cribbing and non-causal states, two states}
		\end{figure}
	\begin{figure}[t]
		\centering
		\psfragscanon
		\psfragfig*[mode=nonstop,scale=0.5]{MAC_cribbing_2states}{
			\psfrag{A}[][][1]{\footnotesize $M_1$}
			\psfrag{B}[][][1]{\footnotesize $M_2$}
			\psfrag{C}[][][1]{\footnotesize $S_1^n$}
			\psfrag{D}[][][1]{\footnotesize Encoder $1$}
			\psfrag{E}[][][1]{\footnotesize Encoder $2$}
			\psfrag{F}[][][1]{\footnotesize $x_{1,i}(M_1,S_1^n)$}
			\psfrag{G}[][][1]{\footnotesize $z(X_{1,i},S_{1,i})$}
			\psfrag{H}[][][1]{\footnotesize $x_{2,i}(M_2,S_2^n,Z^{i})$}
			\psfrag{I}[][][1]{\footnotesize $p_{Y|X_1,X_2,S_1,S_2}$}
			\psfrag{J}[][][1]{\footnotesize $Y_i$}
			\psfrag{K}[][][1]{\footnotesize $(S_1^n,S_2^n)$}			
			\psfrag{L}[][][1]{\footnotesize Decoder}
			\psfrag{M}[][][1]{\footnotesize $(\hat{M}_1,\hat{M}_2)$}	
			\psfrag{O}[][][1]{\footnotesize $S_2^n$}
		}
		\psfragscanoff
		\caption{State dependent MAC with two state components and one side cribbing. The cribbing is causal -- $X_2=x_{2,i}(M_2,S_2^n,Z^{i})$.}
		\label{Figure: DM-MAC with one side cribbing and non-causal states, two states, causal}
	\end{figure}
		\begin{definition}[Code for MAC]\label{Definition: code for MAC with two states}
			A $(R_1,R_2,n)$ code $\mathcal{C}_n$ for the state-dependent MAC with \emph{strictly} causal partial cribbing and two state components is defined by 
			\begin{align*}
				x_{1}^n&:[1:2^{nR_1}]\times S_1^n \to \mathcal{X}_1^n\\
				x_{2,i}&:[1:2^{nR_2}]\times \mathcal{S}_2^n \times Z^{i-1} \to \mathcal{X}_2&1\leq i \leq n \\
				\hat{m}_1&:\mathcal{Y}^n\times \mathcal{S}_1^n \times \mathcal{S}_2^n \to [1:2^{nR_1}] \\
				\hat{m}_2&:\mathcal{Y}^n\times \mathcal{S}_1^n \times \mathcal{S}_2^n \to [1:2^{nR_2}]
			\end{align*}
			for any $\epsilon>0$ and some sufficiently large $n$. 
		\end{definition}
		\begin{definition}[Achievable rate-pair]
			A rate-pair $(R_1,R_2)$ is achievable if there exists a code $\mathcal{C}_n$ such that 
			\begin{align*}
				P_e(\mathcal{C}_n)\defeq\Pr_{\mathcal{C}_n} \left[(\hat{m}_1(Y^n,S_1^n,S_2^n),\hat{m}_2(Y^n,S_1^n,S_2^n))\neq (M_1,M_2)\right] \leq \epsilon
			\end{align*}
			for any $\epsilon>0$ and some sufficiently large $n$.
		\end{definition}
		The capacity region of this channel is defined to be the union of all achievable rate-pairs. 
		We note here that a setup with \emph{causal} cribbing, depicted in Fig. \ref{Figure: DM-MAC with one side cribbing and non-causal states, two states, causal}, satisfy a similar definition with $x_{2,i}:[1:2^{nR_2}]\times \mathcal{S}_2^n \times Z^{i} \to \mathcal{X}_2$.
		\begin{theorem}\label{Theorem: MAC One Side Cribbing non-causal Capacity, two components}
			The capacity region for discrete memoryless MAC with non-causal CSI and \emph{strictly} causal cribbing in Fig. \ref{Figure: DM-MAC with one side cribbing and non-causal states, two states} is given by the set of rate pairs $(R_1,R_2)$ that satisfy
			\begin{subequations}\label{Equation: MAC non-causal capacity region, two states components}
				\begin{align}
				R_1 &\leq I(X_1;Y|X_2,Z,S_1,S_2,U) + H(Z|S_1,U) - I(U;S_1|S_2) \\
				R_2 &\leq I(X_2;Y|X_1,S_1,S_2,U) \\
				R_1+R_2 &\leq I(X_1,X_2;Y|Z,S_1,S_2,U) + H(Z|S_1,U) - I(U;S_1|S_2) \\
				R_1+R_2 & \leq I(X_1,X_2;Y|S_1,S_2) 
				\end{align}
				for PMFs of the form $p_{X_1,U|S_1}p_{X_2|U,S_2}$, with $Z=z(X_1,S_1)$, that satisfies
				\begin{align}
				I(U;S_1|S_2) &\leq H(Z|S_1,U),
				\end{align}
			\end{subequations}
			and $\lvert \mathcal{U} \rvert \leq \min\left\{|\mathcal{S}_2||\mathcal{S}_1||\mathcal{X}_1| |\mathcal{X}_2|+2,|\mathcal{S}_1||\mathcal{S}_2||\mathcal{Y}|+3\right\}$.
		\end{theorem}
		\begin{theorem}\label{Theorem: MAC One Side Cribbing non-causal Capacity, two components, causal}
			The capacity region for discrete memoryless MAC with non-causal CSI and \emph{causal} cribbing in Fig. \ref{Figure: DM-MAC with one side cribbing and non-causal states, two states, causal} is given by the set of rate pairs $(R_1,R_2)$ that satisfy
			the equations in \eqref{Equation: MAC non-causal capacity region, two states components} for PMFs of the form $p_{X_1,U|S_1}p_{X_2|Z,U,S_2}$.
		\end{theorem}
		We note here that when $S_2$ is degenerated, i.e., there is only one state component, the capacity region in both theorems is given by degenerating $S_2$. Note that the difference between Theorems \ref{Theorem: MAC One Side Cribbing non-causal Capacity, two components} and \ref{Theorem: MAC One Side Cribbing non-causal Capacity, two components, causal} is conditioning on $Z$ in the PMF $p_{X_2|Z,U,S_2}$. 
		Here, the auxiliary random variable $U$ plays a double role.
		The first role is similar to the role in the SD-RC; it creates dependency between $X_2$ and $S_1$. 
		This is done using a cooperation codeword $U^n$; Encoder 1 selects a codeword that is coordinated with the states. 
		Encoder 2 uses this codeword in order to cooperate. Since the codeword depends on the state, so does $X_2^n$. 
		When there are two state components, the second component is used by Encoder $2$ to select the cooperation codeword from a collection.  
		The second role is to generate a common message between the encoders. 
		\par In Section \ref{Section: Proof MAC with one state} we provide proof for Theorem \ref{Theorem: MAC One Side Cribbing non-causal Capacity, two components} when there is only one state component. The proof for the general case is given in Section \ref{Section: Proof for MAC with two states} and is based on the case with a single state component.
		The proof for Theorem \ref{Theorem: MAC One Side Cribbing non-causal Capacity, two components, causal} is given in Section \ref{Section: Proof for MAC with two states, causal}. In the following section we examine the results in cases which emphasize the role of $U$.
	\section{Special Cases}\label{Section: Special cases}
	
	\subsection{Cases of State-Dependent SD-RC}
	\par \emph{Case 1: SD-RC without states:}
	When there is no state to the channel, i.e., the channel is fixed throughout the transmission, the capacity of SD-RC is given by Cover and El-Gamal \cite{el1982sdrc} as
	\begin{align}\label{Equation: SD-RC without states capacity}
		\max_{p_{X_r,X}} \min\left\{I(X,X_r;Y),I(X;Y|X_r,Z) + H(Z|X_r)\right\}.
	\end{align}
	This case is captured by degenerating $S$. Then, $S$ can be omitted from the information terms in Theorem \ref{Theorem: SD-RC non-causal capacity} and the joint PMF is $p_{U}p_{X_r|U}p_{X|U}1_{Z|X,X_r}p_{Y|X,X_r}$. Choosing $U=X_r$ recovers the capacity.
	Therefore, we see that here, $U$ plays the role of a common message between $X_r$ and $X$.
	\par \emph{Case 2: SD-RC with causal states}
	Consider a similar configuration to that in Fig. \ref{Figure: SD-RC with non-causal states}, and assume that the states are known to the encoder in a \emph{causal} manner. Although this is not a special case of the non-causal configuration, it emphasizes the role of $U$ further.
	The capacity for this channel was characterized by Kolte \emph{el al} \cite[Theorem 2]{kolte2016binning} by
	\begin{align}\label{Equation: Capacity of SD-RC with causal states}
		C=\max_{p_{X_r}p_{X|X_r,S}} \min\left\{I(X,X_r;Y|S),I(X;Y|X_r,Z,S) + H(Z|X_r,S)\right\}
	\end{align}
	where $Z=z(X,X_r,S)$. Let us compare this capacity to the one with non-causal states. In the causal case, we see that $X$ and $X_r$ are dependent, but $X_r$ and $S$ are not. In the non-causal case (eq. \eqref{Equation: SD-RC capacity non-causal}), $X_r$ and $S$ are dependent. The random variable $U$ generates empirical coordination w.r.t. $P_{U|S}$, and then uses it as common side information at the encoder, relay and decoder. 
	When the state is known causally, such dependency cannot be achieved since the the states are drawn i.i.d. and the relay observes only past outputs of the channel. The capacity of the causal case is directly achievable by Theorem \ref{Theorem: SD-RC non-causal capacity} by substituting $U=X_r$ and $X_r \independent S$.
	\subsection{Cases of State-Dependent MAC with Partial Cribbing}
	Let us investigate the role of the auxiliary random variable $U$ in the MAC configuration via special cases of Theorem \ref{Theorem: MAC One Side Cribbing non-causal Capacity, two components}. We consider here the naive case of one state component, i.e., $S_2$ is degenerated. We denote $S\defeq S_1$ to emphasize this. Proofs for these cases are given in Appendix \ref{Appendix: proofs for special cases of MAC}.
	\par \emph{Case A: Multiple Access Channel with states (without cribbing):}\label{Special cases: MAC as MAC with states}
	\begin{figure}[h!]
		\centering
		\psfragscanon
		\psfragfig*[mode=nonstop,scale=0.5]{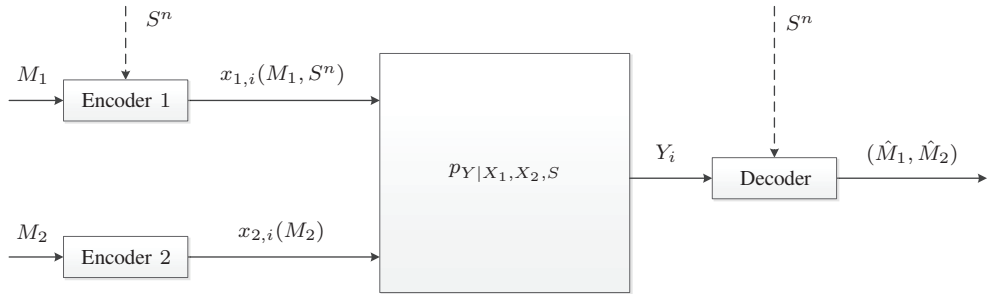}{
			\psfrag{A}[][][1]{\footnotesize $M_1$}
			\psfrag{B}[][][1]{\footnotesize $M_2$}
			\psfrag{C}[][][1]{\footnotesize $S^n$}
			\psfrag{D}[][][1]{\footnotesize Encoder $1$}
			\psfrag{E}[][][1]{\footnotesize Encoder $2$}
			\psfrag{F}[][][1]{\footnotesize $x_{1,i}(M_1,S^n)$}
			\psfrag{H}[][][1]{\footnotesize $x_{2,i}(M_2)$}
			\psfrag{I}[][][1]{\footnotesize $p_{Y|X_1,X_2,S}$}
			\psfrag{J}[][][1]{\footnotesize $Y_i$}
			\psfrag{K}[][][1]{\footnotesize $S^n$}			
			\psfrag{L}[][][1]{\footnotesize Decoder}
			\psfrag{M}[][][1]{\footnotesize $(\hat{M}_1,\hat{M}_2)$}	
		}
		\psfragscanoff
		\caption{Case A - MAC with CSI at one encoder.}
		\label{Special cases: Figure - MAC w CSI}
	\end{figure}
	Consider the case of a multiple access channel with CSI at Encoder 1 and the decoder, depicted in Fig. \ref{Special cases: Figure - MAC w CSI}. It is a special case without cribbing (i.e. $z=\text{constant}$). The capacity region, characterized by Jafar \cite{jafar2006mac}, is defined by all $(R_1,R_2)$ pairs that satisfy
	\begin{subequations}\label{Equation: MAC with STATES speical case, region 1}
		\begin{align}
		R_1\leq & I(X_1;Y|X_2,S) \\
		R_2\leq & I(X_2;Y|X_1,S) \\
		R_1+R_2 \leq & I(X_1,X_2;Y|S)
		\end{align}
	\end{subequations}
	with PMFs that factorize as $p_{X_1|S}p_{X_2}$. 
	\par \emph{Case B: Multiple Access Channel with Conferencing:}\label{Speical cases: MAC as MAC with cooperation}
	\begin{figure}[h!]
		\centering
		\psfragscanon
		\psfragfig*[mode=nonstop,scale=0.5]{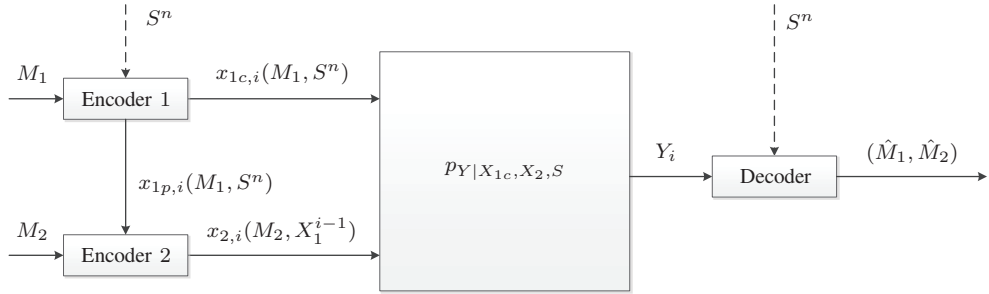}{
			\psfrag{A}[][][1]{\footnotesize $M_1$}
			\psfrag{B}[][][1]{\footnotesize $M_2$}
			\psfrag{C}[][][1]{\footnotesize $S^n$}
			\psfrag{D}[][][1]{\footnotesize Encoder $1$}
			\psfrag{E}[][][1]{\footnotesize Encoder $2$}
			\psfrag{F}[][][1]{\footnotesize $x_{1c,i}(M_1,S^n)$}
			\psfrag{G}[][][1]{\footnotesize \hspace*{3mm} $x_{1p,i}(M_1,S^n)$}
			\psfrag{H}[][][1]{\footnotesize $x_{2,i}(M_2,X_1^{i-1})$}
			\psfrag{I}[][][1]{\footnotesize $p_{Y|X_{1c},X_2,S}$}
			\psfrag{J}[][][1]{\footnotesize $Y_i$}
			\psfrag{K}[][][1]{\footnotesize $S^n$}			
			\psfrag{L}[][][1]{\footnotesize Decoder}
			\psfrag{M}[][][1]{\footnotesize $(\hat{M}_1,\hat{M}_2)$}	
		}
		\psfragscanoff
		\caption{Case B - MAC with CSI at one encoder and conferencing.}
		\label{Special cases: Figure - MAC w Conference}
	\end{figure}
	Consider a case of MAC with conferencing, as depicted in Fig. \ref{Special cases: Figure - MAC w Conference}. In this case, the channel depends only on part of $x_1$, which we denote by $x_{1c}$. The other part of $x_1$, denoted by $x_{1p}$, is known in a strictly causal manner to Encoder 2.  
	\par This setting is different from previous works, which considered a rate-limited cooperation. Here we use a sequence with noiseless communication and a fixed alphabet $\mathcal{X}_{1p}$. It turns out that the capacity region of the channel is the same for both a strictly causal and a non-causal cooperation link. The capacity of both cases when $X_{2,i}=x_{2,i}(M_2,X_{1p}^{i-1})$ and $X_{2,i}=x_{2,i}(M_2,X_{1p}^n)$ is
	\begin{subequations}\label{Equation: MAC with cooperation, region 1}
		\begin{align}
		R_1 	\leq 	& I(X_{1c};Y|U,S) + R_{12} - I(U;S) \\
		R_2 	\leq 	& I(X_2;Y|X_{1c},U,S) \\
		R_1+R_2 \leq 	& \min\left\{I(X_{1c},X_2;Y),I(X_{1c},X_2;Y|U) + R_{12} - I(U;S)\right\} \\
		R_{12} 	=		& \log_2|\mathcal{X}_{1p}|
		\end{align}
	\end{subequations}
	for $p_{U,X_{1c}|S}p_{X_2|U}p_{Y|X_{1c},X_2,S}$.
	\par \emph{Case C: Point-to-point with non-causal CSI:}\label{Special cases: MAC as PTP with states}
	\begin{figure}[h!]
		\centering
		\psfragscanon
		\psfragfig*[mode=nonstop,scale=0.5]{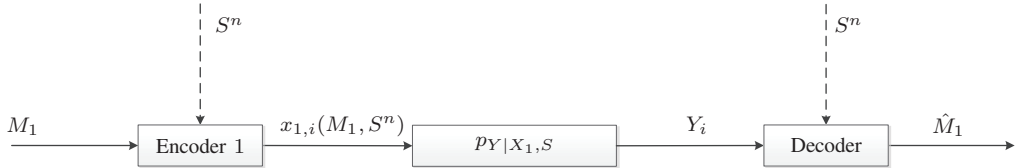}{
			\psfrag{A}[][][1]{\footnotesize $M_1$}
			\psfrag{B}[][][1]{\footnotesize $S^n$}
			\psfrag{C}[][][1]{\footnotesize Encoder $1$}
			\psfrag{D}[][][1]{\footnotesize $x_{1,i}(M_1,S^n)$}
			\psfrag{E}[][][1]{\footnotesize $p_{Y|X_1,S}$}
			\psfrag{F}[][][1]{\footnotesize $Y_i$}
			\psfrag{G}[][][1]{\footnotesize $S^n$}			
			\psfrag{H}[][][1]{\footnotesize Decoder}
			\psfrag{I}[][][1]{\footnotesize $\hat{M}_1$}		
		}
		\psfragscanoff
		\caption{Case C - PTP with non-causal CSI.}
		\label{Special cases: Figure - PTP w CSI}
	\end{figure}
	Consider a configuration of a PTP channel with non-causal CSI, depicted in Fig. \ref{Special cases: Figure - PTP w CSI}. This is a special case of the MAC when $R_2=0$ and $p_{Y_2|X_1,X_2,S}=p_{Y_2|X_1,S}$.
	The capacity of this channel was given by Wolfowitz \cite[Theorem 4.6.1]{wolfowitzBOOK} as 
	\begin{align}\label{Equation: Case B capacity}
	C=\max_{p_{X_1|S}} I(X_1;Y|S).
	\end{align} 
	
\section{Point-to-point with State Encoder and Causality Constraint}\label{Section: PTP w SE}
	\begin{figure}[h!]
		\centering	
		\begin{subfigure}{.45\linewidth}{
				\psfragscanon
				\psfragfig*[mode=nonstop,scale=0.28]{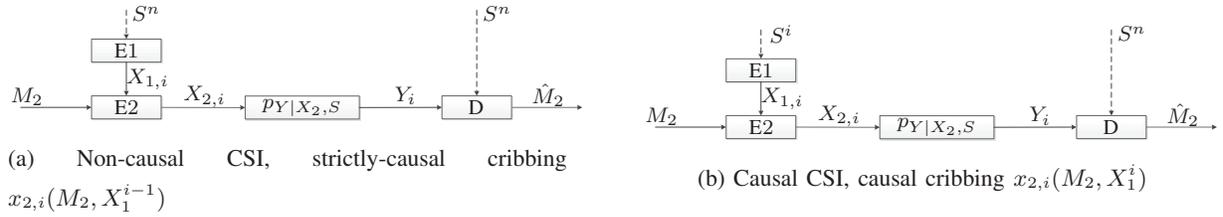}{
					\psfrag{A}[][][1]{\footnotesize $M_2$}
					\psfrag{B}[][][1]{\footnotesize $S^n$}
					\psfrag{C}[][][1]{\footnotesize E$1$}
					\psfrag{D}[][][1]{\footnotesize $X_{1,i}$}
					\psfrag{E}[][][1]{\footnotesize E$2$}
					\psfrag{F}[][][1]{\footnotesize $X_{2,i}$}
					\psfrag{G}[][][1]{\footnotesize $p_{Y|X_2,S}$}
					\psfrag{H}[][][1]{\footnotesize $Y_i$}
					\psfrag{I}[][][1]{\footnotesize $S^n$}			
					\psfrag{J}[][][1]{\footnotesize D}
					\psfrag{K}[][][1]{\footnotesize $\hat{M}_2$}		
				}
				\psfragscanoff
				\caption{Non-causal CSI, strictly-causal cribbing $x_{2,i}(M_2,X_1^{i-1})$}
				\label{Figure: DM-MAC Comparison non-causal and causal 1}
			}
		\end{subfigure}
		\hspace{0.3 in}
		\begin{subfigure}{.45\linewidth}{
				\centering
				\psfragscanon
				\psfragfig*[mode=nonstop,scale=0.28]{PTP_states_both_coded_enc}{
					\psfrag{A}[][][1]{\footnotesize $M_2$}
					\psfrag{B}[][][1]{\footnotesize $S^i$}
					\psfrag{C}[][][1]{\footnotesize E$1$}
					\psfrag{D}[][][1]{\footnotesize $X_{1,i}$}
					\psfrag{E}[][][1]{\footnotesize E$2$}
					\psfrag{F}[][][1]{\footnotesize $X_{2,i}$}
					\psfrag{G}[][][1]{\footnotesize $p_{Y|X_2,S}$}
					\psfrag{H}[][][1]{\footnotesize $Y_i$}
					\psfrag{I}[][][1]{\footnotesize $S^n$}			
					\psfrag{J}[][][1]{\footnotesize D}
					\psfrag{K}[][][1]{\footnotesize $\hat{M}_2$}				
				}
				\psfragscanoff
				\caption{Causal CSI, causal cribbing $x_{2,i}(M_2,X_1^{i})$}
				\label{Figure: DM-MAC Comparison non-causal and causal 2}
			}
		\end{subfigure}
		\caption{Comparison between causal and non-causal CSI.}
		\label{Figure: DM-MAC Comparison non-causal and causal}
	\end{figure}
	\subsection{The State Encoder with a Causality Constraint}
	\par We introduce a new setting, depicted in Fig \ref{Figure: DM-MAC Comparison non-causal and causal}, of a PTP channel with a state encoder (SE) and a causality constraint. 
	The SE has non-causal access to CSI and assists the encoder to increase the transmission rate. 
	The causality constraint enforces the encoder to depend on past observations of the SE.
	This setting is attractive since it is a special case of the MAC, and similar settings may be special cases of more complicated models.	
	\par The setting is defined for two cases; one with non-causal CSI and the other with causal CSI. Explicitly, the setting with non-causal CSI is defined by a state encoder (E1) $x_{1,i}:\mathcal{S}^n\to \mathcal{X}_1$, an encoder (E2) $x_{2,i}:[1:2^{nR_2}]\times \mathcal{X}_1^{i-1}\to \mathcal{X}_2$ and a decoder (D). 
	Note that the encoder depends on strictly causal information from the state encoder.
	The second setting, however, is defined slightly different. 
	First, the state encoder depends on causal CSI, i.e., $x_{1,i}:\mathcal{S}^i\to \mathcal{X}_1$. 
	Secondly, the encoder can use causal information from the state encoder and not strictly causal. Namely, $x_{2,i}:[1:2^{nR_2}]\times \mathcal{X}_1^{i}\to \mathcal{X}_2$.
	We will first discuss on the inclusion of the non-causal case in the MAC setting.
	\par To apply the MAC with partial cribbing to this case, consider the following situation with only one state component. 	
	Encoder 1 has no access to the channel, i.e., $p_{Y|X_1,X_2,S}=p_{Y|X_2,S}$), and no message to send ($R_1=0$). Its only job is to assist Encoder 2 by compressing the CSI and sending it via a private link. The private link is the partial cribbing with $z(x_1,s)=x_1$.
	When the link between the encoders is non-causal, i.e., when $x_{2,i}=f(M_2,X_1^n)$,  using the characterization of Rosenzweig \cite{rosenweig2005partialcsi} with a rate limit of $R_s = \log |\mathcal{X}_1|$ yields
	\begin{align}\label{Equation: State encoder capacity}
	C=\max_{\substack{p_{U|S}p_{X_2|U}: \\I(U;S) \leq \log_2|\mathcal{X}_1|\\}} I(X_2;Y|U,S).
	\end{align}
	When there is a causality constraint, the transmission at time $i$ can only depend on the strictly causal output of state encoder, i.e., $x_{2,i}=f(M_2,X_1^{i-1})$; nonetheless, the capacity remains.
	\par Briefly explained, the capacity is achieved as follows. The transmission is divided to blocks (block-Markov coding). In each block, Encoder 1, which serves as the state encoder, sends a compressed version of the states of the next block. After each transmission block, Encoder $2$ has a compressed version of the state of the current transmission block and uses it for coherent transmission.
	\subsection{An Example - Non-causal CSI Increases Capacity}
	\par The non-causal CSI in the MAC configuration does increase the capacity region in the general case. The following example proves this claim. Consider a model where the channel states are coded, as depicted in Fig. \ref{Figure: DM-MAC Comparison non-causal and causal}. Case (a) is a non-causal case, and (b) is causal. As we previously discussed, the channel in Fig. \ref{Figure: DM-MAC Comparison non-causal and causal 1} is a special case of the \emph{non-causal} state dependent MAC with partial cribbing. Similarly, Fig. \ref{Figure: DM-MAC Comparison non-causal and causal 2} is a special case of \emph{causal} state dependent MAC with partial cribbing \cite{kolte2016binning}.
	\par Since this is a point-to-point configuration, it is a bit surprising that the non-causal CSI increases capacity; when the states are perfectly provided to the encoder, the capacity with causal CSI and with non-causal CSI coincide. As we will next show, in the causal case, the size of $\mathcal{X}_1$ can enforce lossy quantization on the state, while in the non-causal case, the states can be losslessly compressed.
	\begin{figure}[h!]
		\centering	
		\begin{subfigure}{0.3 \linewidth}{
				\centering
				\psfragscanon
				\psfragfig*[mode=nonstop,scale=0.5]{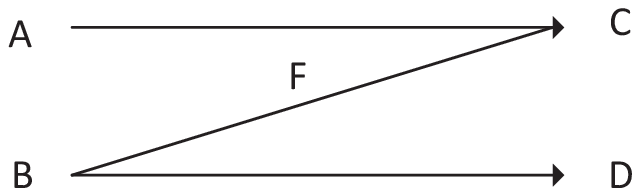}{
					\psfrag{A}[][][1]{\footnotesize $0$}
					\psfrag{B}[][][1]{\footnotesize $1$}
					\psfrag{C}[][][1]{\footnotesize $0$}
					\psfrag{D}[][][1]{\footnotesize $1$}	
					\psfrag{F}[][][1]{\footnotesize $\alpha$}	
				}
				\psfragscanoff
				\caption*{$S=0$}
				\label{Figure: Example - Channel s0}
			}
		\end{subfigure}
		\hspace{0.05 in}
		\begin{subfigure}{0.3 \linewidth}{
				\centering
				\psfragscanon
				\psfragfig*[mode=nonstop,scale=0.5]{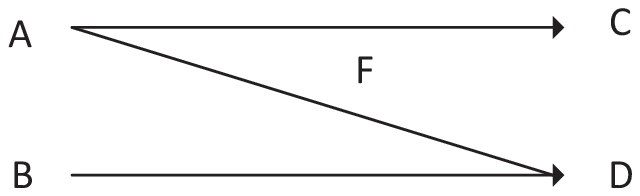}{
					\psfrag{A}[][][1]{\footnotesize $0$}
					\psfrag{B}[][][1]{\footnotesize $1$}
					\psfrag{C}[][][1]{\footnotesize $0$}
					\psfrag{D}[][][1]{\footnotesize $1$}
					\psfrag{F}[][][1]{\footnotesize $\alpha$}
				}
				\psfragscanoff
				\caption*{$S=1$}
				\label{Figure: Example - Channel s1}
			}
		\end{subfigure}
		\hspace{0.05 in}
		\begin{subfigure}{0.3 \linewidth}{
				\centering
				\psfragscanon
				\psfragfig*[mode=nonstop,scale=0.5]{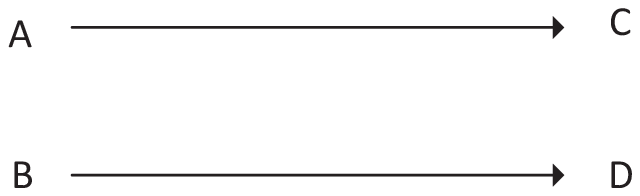}{
					\psfrag{A}[][][1]{\footnotesize $0$}
					\psfrag{B}[][][1]{\footnotesize $1$}
					\psfrag{C}[][][1]{\footnotesize $0$}
					\psfrag{D}[][][1]{\footnotesize $1$}
				}
				\psfragscanoff
				\caption*{$S=2$}
				\label{Figure: Example - Channel s2}
			}
		\end{subfigure}
		\caption{Example of a state dependent channel.}
		\label{Figure: Example - Channel}
	\end{figure}
	\par For every channel $p_{Y|X_2,S}$ and states distribution $p_S$, 
	\begin{align}\label{Example: Capacity of non-causal}
	C_{\text{nc}} = \max_{\substack{p_{U|S}P_{X_2|U}: \\ I(U;S) \leq \log_2 |\mathcal{X}_1|}} I(X_2;Y|S,U) , \qquad C_{\text{c}} = \max_{1_{X_1|S}p_{X_2|X_1}} I(X_2;Y|S,X_1)
	\end{align}
	where $C_{\text{nc}}$ and $C_{\text{c}}$ are the capacity of non-causal and causal CSI configurations, respectively.
	Assume that the states distribution is
	\begin{align}
	p_S(s) = \begin{cases}
	\frac{p}{2} & \text{if } s=0,1 \\
	1-p 		& \text{if } s=2.
	\end{cases}
	\end{align}
	For each state there is a different channel; these channels are depicted in Fig. \ref{Figure: Example - Channel}; a Z-channel for $s=0$, an S-channel for $s=1$, where both share the same parameter $\alpha$, and a noiseless channel for $s=2$. 
	\par The idea is that when the CSI is known non-causally we can compress $S^n$ while in a causal case we cannot.
	Assume that $X_1$ is binary, and $p$ is small enough, for instance $p=0.2$,  such that
	\begin{align}
	H(S) < \log_2|\mathcal{X}_1| = 1.
	\end{align}
	Therefore, taking $U=S$ satisfies $I(U;S) = H(S) \leq 1$ and results in the non-causal capacity
	\begin{align}
	C_{\text{nc}} 	= & \frac{p}{2}\left(C_{\text{Z-channel}}(\alpha)+C_{\text{S-channel}}(\alpha)\right) + (1-p)
	\end{align}
	where
	\begin{align}
	C_{\text{Z-channel}}(\alpha)=C_{\text{S-channel}}(\alpha) = & H_b\left(\frac{2^{H_b(\alpha) / \bar{\alpha}}}{1+2^{H_b(\alpha) / \bar{\alpha}}}\right) - \frac{{H_b(\alpha) / \bar{\alpha}}}{1+2^{H_b(\alpha) / \bar{\alpha}}} .
	\end{align}
	On the other hand, the capacity for causal CSI is
	\begin{align}
	C_{\text{c}} = \max_{\beta}\left[ \frac{p}{2} C_{\text{Z-channel}}(\alpha) + \frac{p}{2}\left(H_b\left(\beta + \bar{\beta}\alpha\right)-\bar{\beta}H_b(\alpha)\right) + (1-p)H_b(\beta)\right].
	\end{align}
	The capacity can be achieve by one of several deterministic functions $x_{1,i}(S^i)$. Each function, maps both $S=2$ and $S=1/0$ to one letter, and $S=0/1$ to the other letter, respectively. Note that this operation causes a lossy quantization of the CSI. 
	For comparison, we also provide the capacity when there is no CSI at the encoder, which is
	\begin{align}
	C_{\text{no-CSI}} = p\left(H_b\left(\frac{1+\alpha}{2}\right)-0.5H_b(\alpha)\right) + (1-p).
	\end{align}
	\begin{table}[!t]
		\renewcommand{\arraystretch}{1.3}
		\caption{Capacity of PTP with coded CSI - numerical evaluations for $p=0.2$.}
		\label{Table: An example - results}
		\centering
		\begin{tabular}{|c||c|c|c|}
			\hline
			$\mathbf{\alpha}$ & \bfseries No-CSI & \bfseries Causal CSI & \bfseries Non-causal CSI \\
			\hline\hline
			$ \mathbf{0} $ 		& $ 1 $ 		&  $ 1 $		& 	$ 1 $ 			\\ \hline
			$ \mathbf{0.5} $ 	& $ 0.8623 $	& $ 0.8633 $	& 	$\mathbf{0.8644}$ 	\\ \hline
			$ \mathbf{1} $ 		& $ 0.8 $ 		& $ 0.8 $ 		& 	$ 0.8 $ 			\\
			\hline
		\end{tabular}
	\end{table}
	\par The capacity of the channels (non-causal, causal, no CSI) for $p=0.2$ are summarized in Table \ref{Table: An example - results}. There are two points where the three channels results in the same capacity. The first is when $\alpha = 0$; in this case, the channel is noiseless for $s=0,1,2$ and the capacity is $1$. There is no need for CSI at the encoder and, therefore, the capacity is the same (among the three cases). The second point is when $\alpha=1$; the channel is stuck at $0$ and stuck at $1$ for $s=0$ and $s=1$, respectively, and noiseless for $s=2$. In this case we can set $P_{X_1}(1)=0.5$ for every $s$ and achieve the capacity. Therefore, the encoder does not use the CSI in those cases. However, for every $\alpha\in(0,1)$, the capacity of the non-causal case is strictly larger than of the others, which confirms that non-causal CSI indeed increases the capacity region.
\section{Proof for Theorem \ref{Theorem: SD-RC non-causal capacity}}\label{Section: SD-RC capacity}
\subsection{Direct}\label{Subsetion: SD-RC capacity direct}
\begin{figure}[!t]
	\centering
	\psfragscanon
		\psfragfig*[mode=nonstop,scale=0.9]{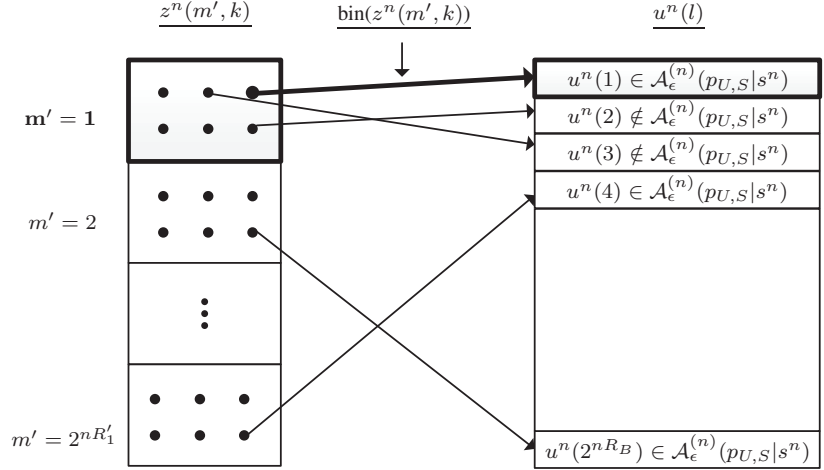}{
		\psfrag{A}[][][1]{\footnotesize $\mathbf{m^\prime=1}$}
		\psfrag{B}[][][1]{\footnotesize $m^\prime=2$}
		\psfrag{C}[][][1]{\footnotesize $m^\prime=2^{nR_1^\prime}$}
		\psfrag{D}[][][1]{\footnotesize $u^n(1)\in\Typical_\epsilon(p_{U,S}|s^n)$}
		\psfrag{E}[][][1]{\footnotesize $u^n(2)\notin\Typical_\epsilon(p_{U,S}|s^n)$}
		\psfrag{F}[][][1]{\footnotesize $u^n(3)\notin\Typical_\epsilon(p_{U,S}|s^n)$}
		\psfrag{G}[][][1]{\footnotesize $u^n(4)\in\Typical_\epsilon(p_{U,S}|s^n)$}
		\psfrag{H}[][][1]{\footnotesize $u^n(2^{nR_B})\in\Typical_\epsilon(p_{U,S}|s^n)$}
		\psfrag{I}[][][1]{\footnotesize \underline{$z^n(m^{\prime},k)$}}
		\psfrag{J}[][][1]{\footnotesize \underline{\text{bin}($z^n(m^{\prime},k)$)}}
		\psfrag{K}[][][1]{\footnotesize \underline{$u^n(l)$}}
	}
	\psfragscanoff
	\caption{Indirect covering: choosing a sequence $z^n$ that points toward a coordinated sequence $u^n$.}
	\label{Figure: Select Bin-Forward}
\end{figure}
\par 
	Before proving the achievability part, let us investigate important properties of the cooperative-bin-forward scheme. This scheme was derived by Kolte \emph{et al} \cite{kolte2016binning} and is based on mapping the discrete finite space $\mathcal{Z}^n$ to a range of indexes $\mathcal{L}=[1:2^{nR_B}]$. We refer this function as cooperative-binning for two reasons: 1) it randomly maps $\mathcal{Z}^n$ into $2^{nR_B}$ bins, and 2) the random binning is independent of all other random variables, which make 'suitable' for cooperation. For instance, a sequence $z^n\in\mathcal{Z}^n$ can be drawn given $v^n$, but its bin index is drawn uniformly, i.e., $\text{bin}(z^n)\sim\text{Unif}[1:2^{nR_B}]$, and is not a function of $v^n$. Thus, if we observe $z^n$ we can find $\text{bin}(z^n)$ without knowing $v^n$. This index is used to create cooperation between the encoder and a relay.
	\begin{lemma}[Indirect covering lemma]\label{Lemma: indirect covering bins}
		Let $\{Z^n(k)\}_{k\in[1:2^{n{R}}]}$ be a collection of sequences, each sequence is drawn i.i.d according to $\prod_{i=1}^{n}p_{Z|V}(z_i|v_i)$. For every $z^n\in\mathcal{Z}^n$, let $\text{Bin}(z^n)\sim\text{Unif}[1:2^{nR_B}]$.
		For any $\delta_1,\delta_2 > 0$, if
		\begin{align}
		{R} <& H(Z|V) - \delta_1 \\
		{R} <& R_B - \delta_2,
		\end{align}
		then,
		\begin{align}
		\lim_{n\to\infty}\Pr\left[\lvert\{ l:\exists k\text{ s.t. Bin}(Z^n(k))=l \}\rvert < 2^{n({R}-\delta_n)}\vert V^n=v^n\right] = 0
		\end{align}
		where $\delta_n\to0$ as $n\to \infty$,
	\end{lemma}
	The proof for this lemma is given in Appendix \ref{Appendix: Covering lemma}.
	Lemma \ref{Lemma: indirect covering bins} states that by choosing ${R}<H(Z|V)-\delta_1$ and $R_B > {R} + \delta_2$, we can guarantee (with high probability) that we will see approximately $2^{n(R-\Delta_n)}$ different bins indexes. Having these indexes allow us to assign to each one of them a sequence or threat them as bins (i.e. use the index to create a list). For instance, if we assign each index $l\in[1:2^{nR_B}]$ a sequence $u^n(l)\sim\prod_{i=1}^{n}p_U(u_i(l))$, we can perform covering \cite[Lemma 3.3]{el2011network} in order to create coordination with another sequence $s^n$, by choosing $R > I(U;S) + \delta$.
	\par The coding scheme works as follows. Divide the transmission to $B$ block and choose a distribution $p_{X,U|S}p_{X_r|S}$. Draw a codebook for each block $b$ which consist of the followings. A cooperative-binning function (a map from  $\mathcal{Z}^n$ to $[1:2^{nR_B}]$, drawn uniformly), a collection of $2^{n(I(U;S)+\delta)}$ codewords $z^{n(b)}(m^{\prime(b)},k)$ for each $m^{\prime(b)}\in[1:2^{nR^\prime}]$ indexed by $k\in[1:2^{n\tilde{R}}]$, a sequence $x^{n(b)}(m^{\dprime(b)})$ for each $m^{\dprime(b)}\in[1:2^{nR^{\dprime}}]$, cooperation codeword $u^{n(b)}(l)$ and a relay codeword  $x_r^{n(b)}(u^{n(b)}(l))$ for each $l\in2^{nR_B}$. 
\par To send a message $m^{(b)}$, recall that the link from the encoder to the relay is deterministic. Therefore, the Encoder can dictate which sequence the relay will observe during the block. Thus, it look at the collection of $z^{n(b)}$ sequences and search for $k$ s.t. $z^{n(b)}(m^{\prime(b)},k)$ points toward a cooperation codeword $u^n$ that is coordinated (typical) with $s^{n(b+1)}$ of the next block. 
This lookup is illustrated in Fig. \ref{Figure: Select Bin-Forward}, and we refer it as \emph{indirect covering}\footnote{For each $z^n$ there is a bin index, and for each bin index there is an $u^n$ sequence. Therefore, the covering is called \emph{indirect}.}. 
Lemma \ref{Lemma: indirect covering bins} guarantees us that if we take $R_B>\tilde{R}>I(U;S)$ and $\tilde{R} < H(Z|X_r,U,S)$ then with high probability we will see at least one coordinated sequence $u^{n(b+1)}$.
Afterwards, the transmission codeword $x^{n(b)}$ is chosen according to $m^{\dprime(b)}$. In the next block, the relay codeword $x_r^{n(b+1)}$ is chosen given $u^{n(b+1)}$. Note that $x_r^{n(b+1)}$ is coordinated with $s^n$ through $u^{n(b+1)}$. 
\par The decoding procedure is done forward using a sliding window technique, derived by Carleial \cite{carleial1982macgeneralized}. At each block $b$, the decoder imitates the encoder procedure for every possible $m^{\prime(b)}\in[1:2^{nR^\prime}]$ and finds $\hat{k}^{(b)}(m^{\prime(b)})$ and $\hat{l}^{(b)}(m^{\prime(b)})$. To ensure that the mapping from $(m^{\prime(b)},\hat{k}^{(b)})$ to $\hat{l}^{(b)}$ is unique, we take $R^\prime + \tilde{R}<R_B$ and $R^\prime +\tilde{R}< H(Z|X_r,U,S)$. Then, the decoder looks for $(\hat{m}^{\prime(b)},\hat{m}^{\dprime(b)})$ such that: 1) all sequences at the current block are coordinated, and 2) $(s^{n(b+1)},u^n(\hat{l}^{(b)}(\hat{m}^{\prime(b)})),x_r^n(u^n(\hat{l}^{(b)})),y^{n(b+1)})$ are coordinated. Setting $R^\dprime < I(X;Y|Z,X_r,U,S)$ and $R < I(X,X_r;Y|S)$ ensures reliability in the decoding procedure.
\par We will now give a formal proof for the achievability part. Fix a PMF $p_{U|S}p_{X_r|U}p_{X|X_r,U,S}$ and let $p_{Z|X_r,U,S}$ be such that $p_{Z|X_r,U,S}p_{X|Z,X_r,U,S}=p_{X|X_r,U,S}1_{Z|X_r,X,S}$.
We use block-Markov coding as follows. Divide the transmission into $B$ blocks, each of length $n$. At each communication block $b$, we transmit a message $M^{(b)}$ at rate $R$. Each message $M^{(b)}$ is divided to $M^{\prime(b)}$ and $M^{\dprime(b)}$, with corresponding rates $R^\prime$ and $R^\dprime$, respectively.
\begin{subequations}
	\par \emph{Codebook:} For each block $b\in[1:B]$, a codebook $\mathcal{C}_n^{(b)}$ is generated as follows:
	\begin{itemize}
		\item[--] \emph{Binning:}  Partition the set $\mathcal{Z}^n$ into $2^{nR_B}$ bins, by choosing uniformly and independently an index $\text{bin}^{(b)}(z^n)\sim U\left[1:2^{nR_B}\right]$.
		\item[--] \emph{Cooperation codewords:} Generate $2^{nR_B}$ $u$-codewords 
		\begin{align}
		&u^n\left(l^{(b-1)}\right)\sim\prod_{i=1}^{n}p_U(u_i), & l^{(b-1)}\in[1:2^{nR_B}]
		\end{align}
		\item[--] \emph{Relay codewords:} 	For each $u^n\in\mathcal{U}^n$ generate $x_r$-codeword $x_r^n\left(u^n\right)\sim\prod_{i=1}^{n}p_{X_r|U}\left(x_{r,i}|u_i\right)$.
		\item[--] \emph{$z$-codewords:} For each $u^n\in\mathcal{U}^n$, $x_r^n\in\mathcal{X}_r^n$ and $s^n\in\mathcal{S}^n$, generate $2^{n(R^\prime+\tilde{R})}$ $z$-codewords 
		\begin{align}
		z^n(m^{\prime(b)},k^{(b)}|x_r^n,u^n,s^{n})\sim\prod_{i=1}^{n}p_{Z|X_r,U,S}(z_i,x_{r,i},s_i),& & m^{\prime(b)}\in[1:2^{nR^\prime}],\; k^{(b)}\in[1:2^{n\tilde{R}}]
		\end{align}
		\item[--] \emph{Transmission codewords:} For each $z^n\in\mathcal{Z}^n$, $u^n\in\mathcal{U}^n$, $x_r^n\in\mathcal{X}_r^n$ and $s^n\in\mathcal{S}^n$ draw $2^{nR^\dprime}$ $x$-codewords 
		\begin{align}
		x^n(m^{\dprime(b)}|z^n,x_r^n,u^n,s^n)\sim\prod_{i=1}^{n}p_{X|Z,X_r,U,S}(x_i|z_i,x_{r,i},u_i,s_i),& & m^{\dprime(b)}\in[1:2^{nR^\dprime}]
		\end{align}
	\end{itemize}
\end{subequations}
\par The block-codebook $\mathcal{C}_n^{(b)}$ consist of all the sequences that was generated for this block. Note that by this construction, all block-codebooks are independent of each other.
\par \emph{Encoder:}	Let $l^{(0)}=m^{\prime(1)}=m^{\dprime(1)}=m^{\prime(B)}=m^{\dprime(B)}=k^{(B)}=1$. This block prefix is done in order to begin the transmission with coordinated cooperation sequence, which is not yet known at the relay. Assume that $l^{(b-1)}$ is known due to former operations at the encoder, and denote 
\begin{align}
z^n(m^{\prime(b)},k^{(b)}|l^{(b-1)},s^{n(b)})=z^n(m^{\prime(b)},k^{(b)}|x_r^n(u^n(l^{(b-1)})),u^n(l^{(b-1)}),s^{n(b)}).
\end{align}
First, the encoder finds $k^{(b)}$ such that
\begin{align}
\left(u^n(\text{bin}(z^n(m^{\prime(b)},k^{(b)}|l^{(b-1)},s^{n(b)}))),s^{n(b+1)}\right)\in\Typical_\epsilon(p_{S,U})
\end{align}
and sets ${l^{(b)} = \text{bin}\left(z^n(m^{\prime(b)},k^{(b)}|l^{(b-1)},s^{n(b)})\right)}$. 
Then, it sends
\begin{align}
x^n\left(m^{\dprime(b)}|z^n(m^{\prime(b)},k^{(b)}|l^{(b-1)},s^{n(b)}),x_r^n(u^n(l^{(b-1)})),u^n(l^{(b-1)}),s^{n(b)}\right).
\end{align}
We abbreviate the notation by ${x^n\left(m^{\dprime(b)}|m^{\prime(b)},k^{(b)},l^{(b-1)},s^{n(b)}\right)}$.
\par \emph{Relay:} Assume $l^{(b-1)}$ is known. At block $b$, send $x_r^n\left(u^n(l^{(b-1)})\right)$. Denote this sequence by $x_r^n\left(l^{(b-1)}\right)$. After the relay observes $z^{n(b)}$, it determines $l^{(b)}=\text{bin}(z^{n(b)})$.
\par \emph{Decoder:}	
\begin{figure*}[t!]
	\begin{subequations}\label{Equation: SD-RC typicality test}
		\begin{align}
		&\left(s^{n(b)},u^n(l^{(b-1)}),x_r^n(l^{(b-1)}),z^n(\hat{m}^{\prime},\hat{k}(\hat{m}^{\prime})|l^{(b-1)},s^{n(b)}),x^n(\hat{m}^{\dprime(b)}|\hat{m}^{\prime},\hat{k}(\hat{m}^{\prime}),l^{(b-1)},s^{n(b)}),y^{n(b)}\right)\\ \nonumber 
		& \in \Typical_\epsilon(p_{S,U,X_r,X,Z,Y})\\
		&\left(s^{n(b+1)},u^n(\hat{l}^{(b)}(m^{\prime(b)})),x_r^n(\hat{l}^{(b)}(m^{\prime(b)})),y^{n(b+1)} \right)\in\Typical_\epsilon(p_{S,U,X_r,Y})
		\end{align}
	\end{subequations}
	\hrule
\end{figure*}
We perform decoding using a sliding window; this is a decoding procedure that decodes from block $1$ to $B-1$, and therefore reduces the delay for recovering message bits at the decoder \footnote{The sliding window technique turns out to be adequate for the relay channel, but not for the MAC.}.  We start at block $2$, since the first cooperation sequence is not necessarily typical with the states at that block. Moreover, since the first message is fixed, the decoder can imitate the encoders operation and find $l^{(1)}$.
\par Assume $l^{(b-1)}$ is known due to previous decoding operations. At block $b$, the decoder performs:
\begin{enumerate}
	\item For each $m^{\prime(b)}$, look for $\hat{k}^{(b)}(m^{\prime(b)},l^{(b-1)},s^{n(b)},s^{n(b+1)})$ and $\hat{l}^{(b)}(m^{\prime(b)},l^{(b-1)},s^{n(b)},s^{n(b+1)})$ the same way that the encoder does. 
	We denote these indexes by $\hat{k}^{(b)}\left(m^{\prime(b)}\right)$ and  $\hat{l}^{(b)}\left(m^{\prime(b)}\right)$. 
	\item Look for unique $(\hat{m}^{\prime},\hat{m}^{\dprime})$ such that \eqref{Equation: SD-RC typicality test} are satisfied.
\end{enumerate}
\par \emph{Analysis of error probability:} 
The code $\mathcal{C}_n$ is defined by the block-codebooks and the encoders and decoder functions.
We bound the average probability of an error at block $b$, conditioned on successful decoding in blocks $\left[1:b-1\right]$. 
Without loss of generality we assume that $M^{\prime(b)}=1$ for each $b\in[1:B]$.
Define the events

\begin{subequations}
	\begin{align}
	\mathrm{E}_1{(b)}&=\left\{\forall k^{(b)}: \left( U^n(\text{Bin}^{(b)}(Z^n(1,k^{(b)}|L^{(b-1)},S^{n(B)}))),S^{n(b+1)}\right) \notin \Typical_\epsilon\left(p_{S,U}\right)\right\}\\
	\mathrm{E}_2(b)&=\left\{\exists m^{\prime(b)}\neq 1 : \text{Bin}^{(b)}(Z^{n}(m^{\prime(b)},k^{(b)}|L^{(b-1)},S^{n(b)}))=L^{(b)},\text{ for some }k^{(b)}\right\} \\
	\mathrm{E}_3{(b)}&=\left\{ \text{Condition }(\ref{Equation: SD-RC typicality test})\text{is not satisfied by }\left(\hat{m}^{\prime(b)},\hat{m}^{\dprime(b)}\right)= \Big(1,1\Big)\right\}\\
	\mathrm{E}_4(b)&=\left\{ \text{Condition } (\ref{Equation: SD-RC typicality test})\text{ is satisfied by some } \left(\hat{m}^{\prime(b)},\hat{m}^{\dprime(b)}\right)\neq \Big(1,1\Big)		 \right\}\\
	\mathrm{E}_5{(b)}&=\left\{\hat{L}^{(b)}=L^{(b)}\right\} \\
	\tilde{E}(b)&= \bigcup_{j=1}^{b}\left\{ \mathrm{E}_1(j)\cup \mathrm{E}_2(j) \cup \mathrm{E}_3(j)\cup \mathrm{E}_4(j)\cup \mathrm{E}_5(j) \right\} \\
	\end{align}
\end{subequations}
The average probability of an error is upper bounded by
\begin{subequations}
	\begin{align}
	P_e^n =& \expectation_{\mathbb{C}_n}\left[ P_e^n\left(\mathbb{C}_n\right)\right] \\
	\leq& \Pr\left[\tilde{E}(B)\right] \\
	\leq & \sum_{b=1}^{B}\Big[\Pr \left[\mathrm{E}_1(b)|\mathrm{E}_5^c(b-1)\right] +\Pr \left[\mathrm{E}_2(b)|\mathrm{E}_5^c(b-1)\right] + \Pr\left[\mathrm{E}_3(b)|\mathrm{E}_5^c(b-1),\mathrm{E}_1^c(b)\right] \\
	&+\Pr\left[\mathrm{E}_4(b)|\mathrm{E}_5^c(b-1),\mathrm{E}_2^c(b),\mathrm{E}_1^c(b)\right] + \Pr\left[\mathrm{E}_5(b)|\mathrm{E}_5^c(b-1),\mathrm{E}_1^c(b),\mathrm{E}_2^c(b),\mathrm{E}_4^c(b)\right]\Big],
	\end{align}
\end{subequations}
where the second inequality follows from union bound and conditioning\footnote{Assume that $\mathrm{A}$ and $\mathrm{B}$ are two events. Then $\Pr\left[\mathrm{A}\cup \mathrm{B}\right] \leq \Pr\left[A\right] + \Pr\left[B|A^{c}\right]$.}.
We will now investigate the probability of each event.
\begin{itemize}
	\item[--] 
	\emph{Event $[\mathrm{E}_1(b)|\mathrm{E}_5{(b-1)}]$:} 	
	By lemma \ref{Lemma: indirect covering bins}, the probability of seeing less than $2^{n(\tilde{R}-\Delta_n)}$ different bins (indexed by $l$) goes to $0$ if $\tilde{R} < H(Z|S,U) - \delta_1$ and $R_B > \tilde{R} + \delta_2$.
	Denote
	\begin{subequations}
		\begin{align}
		\mathrm{A} =& \left\{\text{there are less than } 2^{n(\tilde{R}-\Delta_n)}\text{ different bin indexes}\right\} \\
		\mathcal{D} =&\left\{l:\;\exists k\; \text{such that Bin}(Z^n(k))=l \right\}
		\end{align}
	\end{subequations}
	Therefore,
	\begin{align}
	\Pr\left[\mathrm{E}_1(b)|{E}_5{(b-1)}\right] \leq& \Pr\left[ \mathrm{E}_1(b)|\mathrm{E}_5{(b-1)},\mathrm{A}^c\right] + \Pr[\mathrm{A}|\mathrm{E}_5(b-1)] \\
	=& \Pr\left[ \forall k,\; \left(U^n(\text{Bin}(Z^n(k))),S^n\right)\in\Typical_\epsilon(P_{U,S}) \bigg|\mathrm{E}_5{(b-1)},\mathrm{A}^c \right] +\epsilon^\prime_n \\
	\leq& \Pr\left[ \bigcap_{l\in\mathcal{D}} \left(U^n(l),S^n\right)\in\Typical_\epsilon(P_{U,S}) \bigg|\mathrm{E}_5{(b-1)},\mathrm{A}^c \right] +\epsilon^\prime_n \\
	\leq& \left(1-2^{-n(I(U;S)+\delta^\prime(\epsilon))}\right)^{2^{n(\tilde{R}-\Delta_n)}} + \epsilon^\prime_n \\
	\leq& \exp\left\{-2^{n(\tilde{R}-I(U;S)-\Delta_n-\delta^\prime(\epsilon))}\right\}+ \epsilon^\prime_n
	\end{align}
	which tend to $0$ when $n\to\infty$ if $\tilde{R} > I(U;S) + \Delta_n + \delta^\prime(\epsilon)$.
	\item[--]
	\emph{Event $\left[\mathrm{E}_2(b)|\mathrm{E}_5^c(b-1)\right]$:} Denote  $Z^{n}\left(m^{\prime(b)},k^{(b)}\right)=Z^{n}\left(m^{\prime(b)},k^{(b)}|L^{(b-1)},S^{n(b)}\right)$. \\
	Consider
	\begin{subequations}
		\begin{align}
		\Pr\big[&\mathrm{E}_2(b)|\mathrm{E}_5^c(b-1)\big] \stackrel{}{=}\Pr\left[\exists m^{\prime(b)}\neq 1 : \text{Bin}^{(b)}(Z^{n}(m^{\prime(b)},k^{(b)}))=L^{(b)},\text{ for some }k^{(b)}\right] \\
		=&\Pr\left[\exists m^{\prime(b)}\neq 1 : \text{Bin}^{(b)}(Z^{n}(m^{\prime(b)},k^{(b)}))=\text{Bin}^{(b)}(Z^n(1,K^{(b)})),\text{ for some }k^{(b)}\right] \\
		\stackrel{(a)}{\leq}&\hspace*{-3mm} \sum_{m^{\prime(b)}>1,k^{(b)}} \hspace*{-3mm}\Pr\left[ \text{Bin}^{(b)}(Z^{n}(m^{\prime(b)},k^{(b)}))=\text{Bin}^{(b)}(Z^n(1,K^{(b)}))\right]\\
		\leq& \hspace*{-3mm}\sum_{m^{\prime(b)}>1,k^{(b)}}\hspace*{-3mm}\Pr\left[
		\text{Bin}^{(b)}(Z^{n}(m^{\prime(b)},k^{(b)}))=\text{Bin}^{(b)}(Z^n(1,K^{(b)})),
		Z^{n}(m^{\prime(b)},k^{(b)})=Z^{n}(1,K^{(b)})
		\right] \\
		&+\hspace*{-6mm}\sum_{m^{\prime(b)}>1,k^{(b)}}\hspace*{-4mm}\Pr\left[
		\text{Bin}^{(b)}(Z^{n}(m^{\prime(b)},k^{(b)}))=\text{Bin}^{(b)}(Z^n(1,K^{(b)})), 
		Z^{n}(m^{\prime(b)},k^{(b)})\neq Z^{n}(1,K^{(b)})
		\right] \\
		\leq& \hspace*{-3mm}\sum_{m^{\prime(b)}>1,k^{(b)}}\hspace*{-3mm} \Pr\left[Z^{n}(m^{\prime(b)},k^{(b)})=Z^{n}(1,K^{(b)})\right] \\
		&+\hspace*{-6mm}\sum_{m^{\prime(b)}>1,k^{(b)}}\hspace*{-4mm}\Pr\left[
		\text{Bin}^{(b)}(Z^{n}(m^{\prime(b)},k^{(b)}))=\text{Bin}^{(b)}(Z^n(1,K^{(b)}))\bigg\vert Z^{n}(m^{\prime(b)},k^{(b)})\neq Z^{n}(1,K^{(b)})\right] \\
		\leq& 2^{n(R^\prime + \tilde{R})}2^{-n(H(Z|X_r,U,S) - \delta_1(\epsilon))} + 2^{n(R^\prime + \tilde{R})}2^{-nR_B}
		\end{align}
	\end{subequations}
	where (a) follows by union bound.
	Therefore, this probability goes to zero if
	\begin{subequations}
		\begin{align}
		R^\prime+\tilde{R} <& R_B \\
		R^\prime+\tilde{R} < & H(Z|X_r,U,S) - \delta_1(\epsilon)
		\end{align}
	\end{subequations}
	\item[--]
	\emph{Event $ \left[\mathrm{E}_3(b)|\mathrm{E}_5^c(b-1),\mathrm{E}_1^c(b)\right] $:} Recall that $(U^n(\hat{L}^{(b)}(1)),S^{n(b)})\in \Typical_\epsilon(p_{S,U})$. Therefore, by the conditional typicality lemma \cite[Chapter 2.5]{el2011network}, the probability of this event goes to zero as $n$ goes to infinity.
	
	\item[--]
	\emph{Event $ \left[\mathrm{E}_4(b)|\mathrm{E}_5^c(b-1),\mathrm{E}_2^c(b),\mathrm{E}_1^c(b)\right] $:} We need to distinct between the events in block $b$ and $b+1$. 
	Note that conditioning on $\mathrm{E}_2^c$ ensures us that for $m^{\prime(b)}\neq 1$ we have $\hat{l}^{(b)}(m^{\prime(b)}) \neq L^{(b)}$. Therefore,  at block $b+1$, for each $m^{\prime(b)}\neq 1$ the tuple $\left(S^{n(b+1)},U^n(\hat{l}^{(b)}(m^{\prime(b)})),X_r^n(\hat{l}^{(b)}(m^{\prime(b)}))\right)$ is independent of $Y^{n(b+1)}$ given $S^{n(b+1)}$.
	Therefore,
	\begin{align}
	\Pr\left[\left(S^{n(b+1)},U^n(\hat{l}^{(b)}(m^{\prime(b)})),X_r^n(\hat{l}^{(b)}(m^{\prime(b)}),Y^{n(b+1)})\right)\in\Typical_\epsilon(p_{S,U,X_r,Y})\right] \leq 2^{-n(I(X_r,U;Y|S)-\delta_3(\epsilon))}
	\end{align}
	At block $b$, $(U^n(\hat{L}^{(b-1)}),X_r^n(\hat{L}^{(b-1)}),S^{n(b)},Y^{n(b)})\in\Typical_\epsilon(p_{S,U,X_r,Y})$ with high probability. Therefore, we consider two cases:
	\begin{enumerate}
		\item $\left(\hat{m}^{\prime(b)},\hat{m}^{\dprime(b)}\right) = \left(>1,*\right)$
		\item $\left(\hat{m}^{\prime(b)},\hat{m}^{\dprime(b)}\right) =  \left(1,>1\right)$
	\end{enumerate}
	\begin{table}[!t]
		\renewcommand{\arraystretch}{1.3}
		\caption{Statistical relations in the decoding procedure for SD-RC.}
		\label{Table: SD-RC statistical relations}
		\centering
		\begin{tabular}{|c|c||c|c|}
			\hline
			\bfseries $\hat{m}^{\prime(b)}$ & \bfseries $\hat{m}^{\dprime(b)}$ & PMF (block $b$)& PMF (block $b+1$) \\
			\hline\hline
			$>1$ & $*$ & $p_{U,Z,X_r,X|S}p_{Y|U,S,X_r}$&$p_{S,U,X_r}p_{Y|S}$\\
			$1$ & $>1$ & $p_{U,Z,X_r,X|S}p_{Y|Z,X_r,U,S}$&$p_{S,U,X_r,Y}$\\
			\hline
		\end{tabular}
	\end{table}
	
	The statistical relations between the chosen sequences (by $\hat{m}^{\prime(b)}$ and $\hat{m^{\dprime(b)}}$) are summarized in Table \ref{Table: SD-RC statistical relations}. A standard application of the packing lemma \cite[Lemma 3.1]{el2011network} derives with the following bounds:
	\begin{subequations}
		\begin{align}
		R <& I(Z,X;Y|X_r,U,S) - \delta_3(\epsilon) + I(X_r,U;Y|S)  - \delta_4(\epsilon) \label{Equation: SD-RC sum-rate packing bound}\\
		R^{\dprime} < & I(X;Y|Z,X_r,U,S) - \delta_5(\epsilon)
		\end{align}
	\end{subequations}
	Note that $I(Z,X;Y|X_r,U,S) +I(X_r,U;Y|S) = I(X_r,X;Y|S)$ since $U\Markov (Z,X_r,X,S) \Markov Y$ form a Markov chain and $Z$ is a function of $(X,X_r,S)$.
\end{itemize}
Following this derivation, the probability of an error goes to zero if
\begin{subequations}\label{Equation: SD-RC rates}
	\begin{align}
	R^\prime+\tilde{R} <& R_B \\
	R^\prime+\tilde{R} < & H(Z|X_r,U,S) - \delta_1(\epsilon) \\
	\tilde{R} > & I(U;S) + \Delta_n + \delta^\prime(\epsilon) \\
	R <& I(X,X_r;Y|S) - \delta_3(\epsilon) - \delta_4(\epsilon) \\
	R^{\dprime} < & I(X;Y|Z,X_r,U,S) - \delta_5(\epsilon)
	\end{align}
\end{subequations}
Performing Fourier-Motzkin elimination (can be done using \cite{gattegno2016fourier}) on the rates in \eqref{Equation: SD-RC rates} yields 
\begin{subequations}
	\begin{align}
	R &\leq I(X,X_r;Y|S) \\
	R &\leq I(X;Y|X_r,Z,S,U) + H(Z|X_r,S,U) - I(U;S)
	\end{align}
\end{subequations} 
Cardinality bounds on the auxiliary random variable $U$ are obtained by performing Convex Cover Method \cite[Appendix C]{el2011network}.
\hfill $\square$
	\subsection{Converse}\label{Subsection: Converse for SD-RC capacity}
	Assume that the rate $R$ is achievable, 
	\begin{align}
	P_e(\mathcal{C}_n) \leq \epsilon.
	\end{align}
	By Fano's inequality, we have
	\begin{align}
	H(M|Y^n,S^n) \leq & H_b(P_e(\mathcal{C}_n)) + P_e(\mathcal{C}_n)\log\left(|\mathcal{M}|-1\right) = n\epsilon_n
	\end{align}
	where
	\begin{align}
	\epsilon_n=\frac{1}{n}H_b(P_e(\mathcal{C}_n)) + P_e(\mathcal{C}_n)R.
	\end{align}
	Note that $\epsilon_n\to0$ when $\epsilon\to0$.
	Consider
	\begin{subequations}
		\begin{align}
		nR = &H(M) \\
		\stackrel{(a)}{=}& H(M|S^n) \\
		\stackrel{(b)}{\leq}& I(M;Y^n|S^n) +n\epsilon_n\\
		=& \sum_{i=1}^{n} I(M;Y_i|Y^{i-1},S^n) + n\epsilon_n \\
		=& \sum_{i=1}^{n} I(M,X^{i};Y_i|Y^{i-1},S^n) + n\epsilon_n \\
		=& \sum_{i=1}^{n} I(M,X^{i};Y_i|Y^{i-1},S^n) + n\epsilon_n \\
		=& \sum_{i=1}^{n} I(M,X^{i},X_{r,i};Y_i|Y^{i-1},S^n) + n\epsilon_n \\		
		\leq& \sum_{i=1}^{n} I(M,X^{i},X_{r,i},Y^{i-1},S^{n\setminus i};Y_i|S_i) + n\epsilon_n \\		
		\stackrel{(c)}{=}& \sum_{i=1}^{n} I(X_{i},X_{r,i};Y_i|S_i) + n\epsilon_n \\		
		\stackrel{(d)}{=}& n\left(I(X_Q,X_{r,Q};Y_Q|S_Q,Q) + \epsilon_n\right)  \\		
		\leq & n\left(I(Q,X_Q,X_{r,Q};Y_Q|S_Q) + \epsilon_n\right)  \\		
		\stackrel{(e)}{=}&n\left(I(X_Q,X_{r,Q};Y_Q|S_Q) + \epsilon_n\right) 
		\end{align}
	\end{subequations}
	where: \\
	(a) - since $M\independent S^n$, \\
	(b) - follows by Fano's inequality, \\
	(c) - since $(M,X^{i-1},Y^{i-1},S^{n\setminus i})\Markov (X_i,X_{r,i},S_i)\Markov Y_i$ is a Markov chain, \\
	(d) - by setting $Q\sim U[1:n]$ to be a time-sharing random variable,\\
	(e) - since $Q \Markov (X_Q,X_{r,Q},S_Q) \Markov Y_Q$ is a Markov chain.\\
	and $S^{n\setminus i} = (S^{i-1},S_{i+1}^n)$.
	
	Next, let $U_i\defeq (S^{i-1},Z^{i-1})$. Consider the identity
	\begin{subequations}
		\begin{align}
		H(Z^n,X_r^n|S^n) = &H(Z^n,X_r^n,S^n) - H(S^n) \\
		\stackrel{(f)}{=} & \sum_{i=1}^{n} \left(H(Z_i,X_{r,i},S_i|Z^{i-1},X_r^{i-1},S^{i-1}) - H(S_i)\right) \\
		\stackrel{}{=} & \sum_{i=1}^{n} \left(H(Z_i|X_{r,i},S_i,Z^{i-1},X_r^{i-1},S^{i-1})+H(X_{r,i},S_i|Z^{i-1},X_r^{i-1},S^{i-1}) - H(S_i)\right) \\
		\stackrel{(g)}{=} & \sum_{i=1}^{n} \left(H(Z_i|X_{r,i},S_i,Z^{i-1},S^{i-1})+H(S_i|Z^{i-1},S^{i-1}) - H(S_i)\right) \\
		\stackrel{}{=} & \sum_{i=1}^{n} \left(H(Z_i|X_{r,i},S_i,U_i)+H(S_i|U_i) - H(S_i)\right) \\
		\stackrel{}{=} & \sum_{i=1}^{n} \left(H(Z_i|X_{r,i},S_i,U_i)- I(U_i;S_i)\right) \\
		\stackrel{(h)}{=} & n\left(H(Z_Q|X_{r,Q},S_Q,U_Q,Q)- I(U_Q;S_Q|Q)\right) \\
		\stackrel{(i)}{=} & n\left(H(Z_Q|X_{r,Q},S_Q,U_Q,Q)- I(U_Q,Q;S_Q)\right) \\
		\end{align}
	\end{subequations}
	where: \\
	(e) - since $S^n$ is i.i.d., \\
	(g) - since $X_r^{i}$ is a function of $Z^{i-1}$, \\
	(h) - by definition of $Q$ as a time-sharing random variable, \\
	(i) - since $Q\independent S_Q$.
	Therefore, the following hold:
	\begin{subequations}
		\begin{align}
		I(U_Q,Q;S_Q) \leq& H(Z_Q|X_{r,Q},S_Q,U_Q,Q) \\
		H(Z^n,X_r^n|S^n) =&  n\left(H(Z_Q|X_{r,Q},S_Q,U_Q,Q)- I(U_Q,Q;S_Q)\right) 
		\end{align}
	\end{subequations}
	The second bound is obtained by
	\begin{subequations}
		\begin{align}
		nR = & H(M) \\
		= & H(M|S^n) \\
		=& H(M,Z^n,X_r^n|S^n) \\
		=& H(Z^n,X_r^n|S^n) + H(M|Z^n,X_r^n,S^n)
		\end{align}
	\end{subequations}
	The second term is upper bounded by
	\begin{subequations}
		\begin{align}
		H(M|Z^n,X_r^n,S^n) \leq & I(M;Y^n|Z^n,X_r^n,S^n) + n\epsilon_n \\
		= & \sum_{i=1}^{n} I(M;Y_i|Y^{i-1},Z^n,X_r^n,S^n) + n\epsilon_n \\
		\leq  & \sum_{i=1}^{n} I(M,Y^{i-1},Z_{i+1}^{n},X_{r}^{n\setminus i},S_{i+1}^{n},X_i;Y_i|Z_i,X_{r,i},S_i,U_i) + n\epsilon_n \\
		\stackrel{(j)}{=} & \sum_{i=1}^{n} I(X_i;Y_i|Z_i,X_{r,i},S_i,U_i) + n\epsilon_n \\
		\stackrel{}{=} & n\left(I(X_Q;Y_Q|Z_Q,X_{r,Q},S_Q,U_Q,Q) + \epsilon_n\right)
		\end{align}
	\end{subequations}
	therefore, 
	\begin{align}
	nR\leq n\left(I(X_Q;Y_Q|Z_Q,X_{r,Q},S_Q,U_Q,Q) + H(Z_Q|X_{r,Q},S_Q,U_Q) - I(Q,U_Q;S_Q) + \epsilon_n\right)
	\end{align}
	where (j) follows since $(M,Y^{i-1},Z_{i+1}^{n},X_{r}^{n\setminus i},S_{i+1}^{n})\Markov (X_i,S_i,U_i,Z_i,X_{r,i}) \Markov Y_i$ is a Markov chain.
	
	We need to show that the following conditions hold:
	\begin{itemize}
		\item 
		$Q$ is independent of $S_Q$, 
		\item 
		The following Markov chains hold
		\begin{subequations}
			\begin{align}
			&(M,X^{i-1},Y^{i-1},S^{n\setminus i})\Markov (X_i,X_{r,i},S_i)\Markov Y_i \\
			&(M,Y^{i-1},Z_{i+1}^{n},X_{r}^{n\setminus i},S_{i+1}^{n})\Markov (X_i,S_i,U_i,Z_i,X_{r,i}) \Markov Y_i
			\end{align}
		\end{subequations}
		\item 
		$p_{Y_Q|X_Q,X_{r,Q},Z_Q,S_Q,U_Q,Q}(y|x,x_r,z,s,u,q) = p_{Y|X,X_{r},Z,S}(y|x,x_r,z,s
		)$
		\item $Z_Q = z(X_Q,X_{r,Q},S_Q)$
	\end{itemize}
	The first condition holds due to the i.i.d. distribution of the states sequence $S^n$.
	The distribution on the random variables is
	\begin{align}
	p(m,&s^n,x^n,x_r^n,z^n,y^n) \nonumber \\ 
	& = p(m)\prod_{i=1}^{n}p(s_i)\prod_{i=1}^{n}1(x_i|m,s^n)1(x_{r,i}|z^{i-1})1(z_i|x_i,x_{r,i},s_i)p_{Y|X,X_r,Z,S}(y_i|x_i,x_{r,i},z_i,s_i).
	\end{align}
	The Markov chains in the second condition can be readily seen from this distribution. Moreover, for each $i$, $Z_i=z(Z_i,Z_{r,i},S_i)$ and the third condition also holds. 
	By defining $U=(U_Q,Q) ,\; X=X_Q, X_r = X_{r,Q}, S= S_Q$ and $Z= Z_Q$, we derive with the following bound:
	\begin{subequations}
		\begin{align}
		R \leq& I(X,X_r;Y|S) +\epsilon_n\\
		R \leq& I(X;Y|X_r,Z,U,S) + H(Z|X_r,S,U) - I(U;S) + \epsilon_n
		\end{align}
		with PMF that factorizes as
		\begin{align}
		p_{U|S}p_{X_r|U}p_{X|X_r,U,S}
		\end{align}
	\end{subequations}
	that satisfies $I(U;S) \leq H(Z|X_r,S,U)$.
	This completes the proof for the converse part. \hfill $\square$
\section{Proof for MAC with one state component}\label{Section: Proof MAC with one state}
	\subsection{Direct}\label{Subsection: MAC achievability for one state}
	\par We first discuss the achievability scheme for the case where $S_2$ is degenerate. To ease the notation we use $S=S_1$. The capacity region for this case is given by 
	\begin{subequations}\label{Equation: MAC non-causal capacity region}
		\begin{align}
		R_1 &\leq I(X_1;Y|X_2,Z,S,U) + H(Z|S,U) - I(U;S) \\
		R_2 &\leq I(X_2;Y|X_1,S,U) \\
		R_1+R_2 &\leq I(X_1,X_2;Y|Z,S,U) + H(Z|S,U) - I(U;S) \\
		R_1+R_2 & \leq I(X_1,X_2;Y|S) 
		\end{align}
		for PMFs of the form $p_{U|S}p_{X_1|S,U}p_{X_2|U}$, with $Z=z(X_1,S)$, that satisfies
		\begin{align}
		I(U;S) &\leq H(Z|S,U),
		\end{align}
	\end{subequations}
	and $|\mathcal{U}| \leq \min\{|\mathcal{S}||\mathcal{X}_1||\mathcal{X}_2|+2,|\mathcal{S}||\mathcal{Y}|+1\}$.
	\par The coding scheme for this case gives the key steps for the general case (in Theorem \ref{Theorem: MAC One Side Cribbing non-causal Capacity, two components}). Note that Encoder 2 plays here a double role: First, it helps Encoder 1 to deliver his message $M_1$ by cribbing $Z^{i-1}$ at each time $i$. This is done using the cooperative-bin-forward scheme as in the SD-RC in section \ref{Section: SD-RC capacity}. Second, it delivers its own message $M_2$ to the decoder using the same transmission sequence $X_2^n$. To do so,  a superposition code is built on the shared common information which is represented by the sequence $U^n$. This common information also coordinated with the states sequence $S^n$. The decoding procedure however is done backward, which is called \emph{backward decoding}.  We will now give a detailed proof for the achievable rates.
	\par Fix a PMF $p_{U|S}p_{X_1|U,S}p_{X_2|U}$ and $\epsilon > 0$. 
	Divide the transmission to $B$ blocks, each of length $n$. At each communication block $b$, we transmit at rates $R_1$ and $R_2$. We perform rate-splitting for $R_1$; at each block $b\in\left[1:B\right]$, split the rate $R_1=R_1^\prime + R_1^{\dprime}$, with the message $M_1^{(b)}=(M_1^{\prime(b)},M_1^{\dprime(b)})$ accordingly. 
	\par \emph{Codebook:} The \textit{codebook} $\mathcal{C}_n$ is defined to be collection of \textit{block-codebooks}, $\left\{\mathcal{C}_n^{(b)}\right\}_{b\in[1:B]}$.
	For each block $b\in[1:B]$, a codebook $\mathcal{C}_n^{(b)}$ is generated as follows:
	\begin{subequations}
	\begin{itemize}
		\item[--] \emph{Binning:} Partition the set $\mathcal{Z}^n$ into $2^{nR_u}$ bins, by drawing uniformly and independently an index \begin{align}
		\text{bin}^{(b)}(z^n)\sim U\left[1:2^{nR_u}\right]&&\forall z^n\in\mathcal{Z}^n
		\end{align}
		\item[--] 
		\emph{Cooperation codewords:} Generate $2^{nR_B}$ $u$-codewords 
		\begin{align}
		u^{n}(l^{(b-1)})\sim \prod_{i=1}^{n}p_{U}(u_i(l^{(b-1)})),&&l^{(b-1)}\in\left[1:2^{nR_B}\right]
		\end{align}
		\item[--] 
		\emph{Cribbed codewords:} For each $u^n\in\mathcal{U}^n$ and $s^n\in\mathcal{S}^n$, generate $2^{n(R_1^\prime + \tilde{R}_1)}$ $z$-codewords, 
		\begin{align}
		z^n(m_1^{\prime(b)},k^{(b)}|u^{n},s^n)\sim\prod_{i=1}^{n}p_{Z|U,S}(z_i|u_i,s_i),&&m_1^{\prime(b)}\in\left[1:2^{nR_1^\prime}\right],\;k^{(b)}\in\left[1:2^{n\tilde{R}_1}\right].
		\end{align}
		\item[--] 
		\emph{Transmission codewords at Encoder 1:} For each $z^n\in\mathcal{Z}^n,\;u^n\in\mathcal{U}^n$ and $s^n\in\mathcal{S}^n$ generate $2^{nR_1^{\dprime}}$ $x_1$-codewords, 
		\begin{align}
		x_1^{n}(m_1^{\dprime(b)}|z^n,u^n,s^n)\sim\prod_{i=1}^{n}p_{X_1|Z,U,S}(x_{1,i}|z_i,u_i,s_i),&&m_1^{\dprime(b)}\in\big[1:2^{nR_1^{\dprime}}\big].
		\end{align}
		\item[--]
		\emph{Transmission codewords at Encoder 2:} For each $u^n\in\mathcal{U}^n$, draw $2^{nR_2}$ $x_2$-codewords, 
		\begin{align}
		x_2^{n}(m_2^{(b)}|u^n)\sim\prod_{i=1}^{n}p_{X_2|U}(x_{2,i}|u_i),&&m_2\in\left[1:2^{nR_2}\right].
		\end{align}
	\end{itemize}
	\end{subequations}
	The block-codebook $\mathcal{C}_n^{(b)}$ consist of all the sequences that was generated for this block. Note that by this construction, all block-codebooks are independent of each other.
	\par \emph{Prefix and suffix blocks:}  Let $m_1^{\prime(1)},m_1^{\dprime(1)},m_1^{\prime(B)},m_1^{\dprime(B)},m_2^{(1)},m_2^{(B)},k^{(B)}$ and $l^{(1)}$ be equal to one.  Namely, at blocks $1$ and $B$ the encoders don't send any message, and hence these blocks are prefix and suffix for the transmission.
	Here, in addition to the suffix block that is used in block Markov coding schemes, the prefix block is used for the encoders to agree on the second cooperation codeword that is typical with $s^{n(2)}$. However, for $l^{(0)}$ the corresponding cooperation sequence $u^n(l^{(0)})$ is not necessarily typical with the states in the first block.
	Due to prefix and suffix blocks, the average rates are $\bar{R}_1 = \frac{B-2}{B}R_1$ and $\bar{R}_2 = \frac{B-2}{B} R_2$. By choosing $B$ sufficiently large, the average rates can be made close to $R_1$ and $R_2$ as desired \footnote{One can show that for every fixed $B$, we can take $n$ to be large enough to make the probability of an error small as desired.}.
	\par \emph{Encoder 1:} Denote by $s^{n(b)}$ the states sequence from block $b$, and
	\begin{align}
		z^n(m_1^{\prime(b)},k^{(b)}|l^{(b-1)},s^{n(b)})\defeq z^n(m_1^{\prime(b)},k^{(b)}|u^n(l^{(b-1)}),s^{n(b)})
	\end{align}
	the cribbed codewords from codebook $\mathcal{C}_n^{(b)}$. 
	At block $b$, the encoder looks for $k^{(b)}$ such that
	\begin{align}
		\left(u^n(\text{bin}^{(b)}(z^n(m_1^{\prime(b)},k^{(b)}|u^n(l^{(b-1)}),s^{n(b)}))),s^{n(b+1)}\right)\in\Typical_\epsilon\left(p_{S,U}\right).
	\end{align}
	If such $k^{(b)}$ cannot be found, choose $k^{(b)}$ uniformly. If more than one was found, choose the first.
	Set 
	\begin{align}
		l^{(b)}=\text{bin}^{(b)}(z^n(m_1^{\prime(b)},k^{(b)}|l^{(b-1)},s^{n(b)}))
	\end{align} and transmits the codeword 
	\begin{align}
		x_1^n(m_1^{\dprime(b)}\big\vert z^n(m_1^{\prime(b)},k^{(b)}|u^n(l^{(b-1)}),s^{n(b)}),u^n(l^{(b-1)}),s^{n(b)})
	\end{align}
	The transmitted codeword is denoted by $x_1^n(m_1^{\dprime(b)}\big\vert m_1^{\prime(b)},k^{(b)},l^{(b-1)},s^{n(b)})$. 
	\par \emph{Encoder 2:} Assuming that $l^{(b-1)}$ is known from previous encoding operations, at block $b$ Encoder 2 transmits $x_2^n(m_2^{(b)}|u^n(l^{(b-1)}))$. We denote this codeword as $x_2^n(m_2^{(b)}|l^{(b-1)})$. 
	At the end of the block, this encoder observes $z^n(m_1^{\prime(b)},k^{(b)}|l^{(b-1)},s^{n(b)})$, and sets $l^{(b)}=\text{bin}^{(b)}(z^n(m_1^{\prime(b)},k^{(b)}|l^{(b-1)},s^{n(b)}))$.
	\begin{figure}[t]
		\begin{align}\label{Equation: DM-MAC decoding typicality}
		\bigg(u^{n}(\hat{l}^{(b-1)}),s^{n(b-1)},z^n(\hat{m}_1^{\prime(b)},\hat{k}^{(b)}|\hat{l}^{(b-1)},&s^{n(b)}),x_1^{n}\big(\hat{m}_1^{\dprime(b)}|\hat{m}_1^{\prime(b)},\hat{k}^{(b)},\hat{l}^{(b-1)},s^{n(b-1)}\big), \dots\nonumber\\
		&x_2^{n}\big(\hat{m}_2^{(b)}|l^{(b-1)}\big),y^{n(b)}\bigg) 	\in \Typical_{2\epsilon}\bigg(p_{S,U,X_1,Z}p_{X_2|U}p_{Y|X_1,X_2,S}\bigg)
		\end{align}
		\hrule
	\end{figure}
	\par \emph{Decoding:} The decoding procedure is done backwards, Laneman and Kramer \footnote{In \cite{laneman2004window} showed that for the MAC (in contrary to SD-RC) sliding window decoding is sometimes inferior to backward decoding in terms of achievable rates.}. We start decoding from block B to block 2.
	Assuming $l^{(b)}$ is known by decoding former blocks, the decoder performs:
	\begin{enumerate}
		\item For each $l^{(b-1)}$, find $\hat{k}^{(b)}\left(l^{(b-1)},l^{(b)},s^{n(b+1)}\right)$ and $\hat{m}_1^{\prime(b)}\left(l^{(b-1)},l^{(b)},s^{n(b+1)}\right)$, abbreviated as $\hat{m}_1^{\prime(b)}(l^{(b-1)})$ and $\hat{k}^{(b)}(l^{(b-1)})$, such that 
		\begin{align}
			\text{bin}(z^{n}(\hat{m}_1^{\prime(b)}(l^{(b-1)}),\hat{k}^{(b)}(l^{(b-1)})|l^{(b-1)},s^{n(b-1)})) = l^{(b)}
		\end{align}
		
		\item Denote the channel's output at blocks $b$ by $y^{n(b)}$. Find a unique tuple $\Big(\hat{l}^{(b-1)},\hat{m}_1^{\dprime(b)},\hat{m}_2^{(b)}\Big)$ such that \eqref{Equation: DM-MAC decoding typicality} is satisfied.
		If such $\left(\hat{l}^{(b-1)},\hat{m}_1^{\dprime(b)},\hat{m}_2^{(b)}\right)$ cannot be found, choose each uniformly.
	\end{enumerate}
	Recall that $m_1^{\prime(B)}=k^{(B)}=m_1^{\dprime(B)}=k^{(B)}=m_2^{(B)}=1$. To initialize the decoding procedure, in block $B$ (first decoding block) find $l^{(B-1)}$ using condition (\ref{Equation: DM-MAC decoding typicality}).
	\par \emph{Analysis of error probability:} The code $\mathcal{C}_n$ is defined by the block-codebooks $\{\mathcal{C}_n^{(b)}\}_{b=1}^B$, the encoders and decoder functions.	We bound the average probability of an error at block $b$; encoding error events are conditioned on successfully encoding blocks $[1:b-1]$, and decoding error events are conditioned on successfully decoding blocks $\left[b+1:B\right]$.
	Without loss of generality we assume that $M_1^{\prime(b)}=M_1^{\dprime(b)}=M_2^{(b)}=1$ for each $b\in[1:B]$.
	Define the events
	\begin{subequations}
		\begin{align}
		\mathrm{E}_1{(b)}&=\left\{\forall k^{(b)}: ( U^n(\text{Bin}^{(b)}(Z^n(1,k^{(b)}|L^{(b-1)},S^{n(B)}))),S^{n(b+1)}) \notin \Typical_\epsilon(p_{S,U})\right\}\\
		\mathrm{E}_2(b)&=\left\{\exists m_1^{\prime(b)}\neq 1 : \text{Bin}^{(b)}(Z^{n}(m_1^{\prime(b)},k^{(b)}|L^{(b-1)},S^{n(b)}))=L^{(b)},\text{ for some }k^{(b)}\right\} \\
		\mathrm{E}_3{(b)}&=\left\{ \text{Condition }(\ref{Equation: DM-MAC decoding typicality})\text{is not satisfied by }(\hat{l}^{(b-1)},\hat{m}_1^{\dprime(b)},\hat{m}_2^{(b)})= \Big(L^{(b-1)},1,1\Big)\right\}\\
		\mathrm{E}_4(b)&=\left\{ \text{Condition } (\ref{Equation: DM-MAC decoding typicality})\text{ is satisfied by some } (\hat{l}^{(b-1)},\hat{m}_1^{\dprime(b)},\hat{m}_2^{(b)})\neq \Big(L^{(b-1)},1,1\Big)		 \right\}\\
		\mathrm{E}_5{(b)}&=\left\{\hat{L}^{(b)}=L^{(b)}\right\} \\
		\tilde{E}_1(b)&= \bigcup_{j=1}^{b}\left\{ \mathrm{E}_1(j)\cup \mathrm{E}_2(j) \right\} \\
		\tilde{E}_2(b)&= \bigcup_{j=b}^{B}\left\{ \mathrm{E}_3(j)\cup \mathrm{E}_4(j)\cup \mathrm{E}_5(j)\right\}
	\end{align}
	\end{subequations}
	The average probability of an error is upper bounded by the union of these events in all blocks,
	\begin{subequations}
		\begin{align}
		P_e^n =& \expectation_{\mathbb{C}_n}\left[ P_e^n\left(\mathcal{C}_n(\mathbb{C}_n)\right)\right] \\
		\leq& \Pr\left[\tilde{E}_1(B)\cup \tilde{E}_2(1)\right] \\
		\leq & \Pr \left[\tilde{E}_1(B)\right] + \Pr \left[\tilde{E}_2(1),\tilde{E}_1^c(B)\right] \\
		\leq & \sum_{b=1}^{B}\Big(\Pr \left[\mathrm{E}_1(b)\right] +\Pr \left[\mathrm{E}_2(b)\right] + \Pr\left[\mathrm{E}_3(b)|\mathrm{E}_5^c{(b+1)},\tilde{E}_1^c(B)\right] \\
		&+
		\Pr \left[\mathrm{E}_4(b)|\mathrm{E}_5^c{(b+1)},\tilde{E}_1^c(B)\right] + \Pr\left[\mathrm{E}_5{(b)}|\mathrm{E}_3^c(b),\mathrm{E}_4^c(b),\mathrm{E}_5^c{(b+1)},\tilde{E}_1^c(B)\right] \Big)
	\end{align}
	\end{subequations}
	\begin{itemize}
		\item[--] \emph{Event $\left[\mathrm{E}_1(b)\right]$:} 	We need to satisfy that the probability of seeing $U^n(l^{(b)})$ that is jointly typical with $S^{n,(b+1)}$ go to $1$ as $n$ goes to infinity. 
		According to Lemma \ref{Lemma: indirect covering bins}, we can ensure that $\Pr\left[\mathrm{E}_1(b)|\tilde{E}_1^c(b-1)\right]\xrightarrow[n\to\infty]{}0$ by taking $R_B > \tilde{R}_1+\delta_1(\epsilon)$, $\tilde{R}_1>I(U;S) + \Delta_n$ and  $\tilde {R}_1 < H(Z|U,S) - \delta_2(\epsilon)$.
		
		\item[--] \emph{Event $\left[\mathrm{E}_2(b)\right]$:} 	Denote  $Z^{n}(m_1^{\prime(b)},k^{(b)})=Z^{n}(m_1^{\prime(b)},k^{(b)}|L^{(b-1)},S^{n(b)})$.
		Consider
		\begin{subequations}
			\begin{align}
			\Pr\big[\mathrm{E}_2(b)&\big] \stackrel{}{=}\Pr\left[\exists m_1^{\prime(b)}\neq 1 : \text{Bin}^{(b)}(Z^{n}(m_1^{\prime(b)},k^{(b)}))=L^{(b)},\text{ for some }k^{(b)}\right] \\
			=&\Pr\left[\exists m_1^{\prime(b)}\neq 1 : \text{Bin}^{(b)}(Z^{n}(m_1^{\prime(b)},k^{(b)}))=\text{Bin}^{(b)}(Z^n(1,K^{(b)})),\text{ for some }k^{(b)}\right] \\
			\leq&\hspace*{-3mm}\sum_{m_1^{\prime(b)}>1,k^{(b)}}\hspace*{-3mm}\Pr\left[
		 \text{Bin}^{(b)}(Z^{n}(m_1^{\prime(b)}, 
		 k^{(b)}))=\text{Bin}^{(b)}(Z^n(1,K^{(b)}))
		 \right]\\
			\leq&\hspace*{-3mm}\sum_{m_1^{\prime(b)}>1,k^{(b)}} \hspace*{-3mm}\Pr\left[
			\text{Bin}^{(b)}(Z^{n}(m_1^{\prime(b)},k^{(b)}))=\text{Bin}^{(b)}(Z^n(1,K^{(b)})),
			Z^{n}(m_1^{\prime(b)},k^{(b)})=Z^{n}(1,K^{(b)})
			\right] \\
			&+ \hspace*{-6mm}\sum_{m_1^{\prime(b)}>1,k^{(b)}}\hspace*{-3mm}\Pr\left[
		 \text{Bin}^{(b)}(Z^{n}(m_1^{\prime(b)},k^{(b)}))=\text{Bin}^{(b)}(Z^n(1,K^{(b)})), 
			Z^{n}(m_1^{\prime(b)},k^{(b)})\neq Z^{n}(1,K^{(b)})
			\right] \\
			\leq&\hspace*{-3mm}\sum_{m_1^{\prime(b)}>1,k^{(b)}} \hspace*{-3mm}\Pr\left[Z^{n}(m_1^{\prime(b)},k^{(b)})=Z^{n}(1,K^{(b)})\right] \\
			&+\hspace*{-6mm}\sum_{m_1^{\prime(b)}>1,k^{(b)}}\hspace*{-3mm}\Pr\left[ \text{Bin}^{(b)}(Z^{n}(m_1^{\prime(b)},k^{(b)}))=\text{Bin}^{(b)}(Z^n(1,K^{(b)}))\bigg\vert Z^{n}(m_1^{\prime(b)},k^{(b)})\neq Z^{n}(1,K^{(b)})\right] \\
			\leq& 2^{n(R_1^\prime + \tilde{R}_1)}2^{-n(H(Z|U,S) - \delta_3(\epsilon))} + 2^{n(R_1^\prime + \tilde{R}_1)}2^{-nR_B}
		\end{align}
		\end{subequations}
		Therefore, this probability goes to zero as $n\to\infty$ if $R_1^\prime + \tilde{R}_1 < H(Z|U,S)-\delta_3(\epsilon)$ and $R_1^\prime + \tilde{R}_1 < R_B$. 
		
		\item[--] \emph{Event $\left[\mathrm{E}_3(b)|\mathrm{E}_5^c{(b+1)},\tilde{E}_1^c(B)\right]$:}
		Note that given $\tilde{E}_1^c(b-1)$, the cooperation codeword $U^n(L^{(b-1)})$ and $S^{n(b)}$ are jointly typical. 
		Recall that by the codebook generated i.i.d, each block-codebook is independent of each other and the channel is memoryless. Thus, by conditional typicality lemma, the probability of this event goes to $0$ when $n\to\infty$.
		
		\item[--] \emph{Event $\left[\mathrm{E}_4(i)|\mathrm{E}_5^c{(b+1)},\tilde{E}_1^c(B)\right]$:} There are several cases in which this error can occur:
		\begin{enumerate}
			\item\label{Cases - Packing lemma, DM-MAC - Case 1} $\left(l^{(b-1)},m_1^{\dprime(b)},m_2^{(b)}\right)=\left(\neq L^{(b-1)}, *, *\right)$
			\item\label{Cases - Packing lemma, DM-MAC - Case 2} $\left(l^{(b-1)},m_1^{\dprime(b)},m_2^{(b)}\right)=\left(L^{(b-1)}, >1, >1\right)$
			\item\label{Cases - Packing lemma, DM-MAC - Case 3} $\left(l^{(b-1)},m_1^{\dprime(b)},m_2^{(b)}\right)=\left(L^{(b-1)}, 1, >1\right)$
			\item\label{Cases - Packing lemma, DM-MAC - Case 4} $\left(l^{(b-1)},m_1^{\dprime(b)},m_2^{(b)}\right)=\left(L^{(b-1)}, >1, 1\right)$
		\end{enumerate}
		\begin{table}[!t]
			\renewcommand{\arraystretch}{1.3}
			\caption{Statistical relations in the decoding procedure for MAC with cribbing}
			\label{Table: MAC statistical relations}
			\centering
			\begin{tabular}{|c|c|c||c|}
				\hline
				\bfseries ${l}^{\prime(b-1)}$ & \bfseries ${m}_1^{\dprime(b)}$ & \bfseries ${m}_2^{(b)}$ & PMF \\
				\hline\hline
				$\neq L^{(b-1)}$ & $*$ & $*$ 	& $p_{S,U,X_1,X_2,Z}p_{Y|S}$\\
				$L^{(b-1)}$ & $>1$ & $>1$ 		& $p_{S,U,X_1,X_2,Z}p_{Y|Z,U,S}$\\
				$L^{(b-1)}$ & $1$ & $>1$ 		& $p_{S,U,X_1,X_2,Z}p_{Y|X_1,Z,U,S}$\\
				$L^{(b-1)}$ & $>1$ & $1$ 		& $p_{S,U,X_1,X_2,Z}p_{Y|X_2,Z,U,S}$\\
				\hline
			\end{tabular}
		\end{table}
		The probability of each case is bounded by standard application of the packing lemma as follows. The statistical relations between the codewords are summarized in Table \ref{Table: MAC statistical relations}.
		Denote by $\tilde{E}_{i,j,k}(b)$ the event that (\ref{Equation: DM-MAC decoding typicality}) is satisfied by $(l^{(b-1)},m_1^{\dprime(b)},m_2^{(b)})=(i, j, k)$. 
		By union-bound, case \ref{Cases - Packing lemma, DM-MAC - Case 1} is upper-bounded by
		\begin{subequations}
			\begin{align}
				\Pr\bigg[&\bigcup_{i\neq L^{(b-1)},j,k}\tilde{E}_{i,j,k}(b)|\mathrm{E}_5^c{(b+1)},\tilde{E}_1(B)\bigg]\leq \sum_{i\neq L^{(b-1)},j,k}\Pr\left[\tilde{E}_{i,j,k}(b)|\mathrm{E}_5^c{(b+1)},\tilde{E}_1(B)\right] \\
				\stackrel{(a)}{\leq}& \hspace*{-5mm}\sum_{i\neq L^{(b-1)},j,k}\hspace*{-3mm}2^{-n(I(U;S) - \delta_4(\epsilon))}\times\Pr\left[\tilde{E}_{i,j,k}(b)|\mathrm{E}_5^c{(b+1)},\tilde{E}_1(B),(U^n(i),S^{n(b)})\in\Typical_{0.5\epsilon}(p_{S,U})\right] \\
				\stackrel{(a)}{\leq}& 2^{n(R_B+R_1^{\dprime}+R_2)}2^{-n(I(U;S)-\delta_4(\epsilon))}2^{-n(I(U,X_1,X_2,Z;Y|S) - \delta_5(\epsilon)}\\
				=& 2^{n(R_B+R_1^{\dprime}+R_2-(I(X_1,X_2;Y|S) - \delta_5(\epsilon)+I(U;S)-\delta_4(\epsilon)}.
			\end{align}
		\end{subequations}
		where: \\
		(a) - follows since for each $i\neq L^{b-1}$, $U^n(i)$ is independent of $S^{n(b)}$, \\
		(b) - follows since for each $i\neq L^{b-1}$,  the codewords corresponding to $(i,j,k)$ are independent of $Y^{n(b)}$ given $S^{n(b)}$. \\
		Similarly, case \ref{Cases - Packing lemma, DM-MAC - Case 2} is upper bounded by 
		\begin{align}
			\sum_{j>1,k>1}\Pr\left[\tilde{E}_{L^{(b-1)},j,k}(b)|\mathrm{E}_5^c{(b+1)},\tilde{E}_1(B)\right] & \leq 2^{n(R_1^{\dprime}+R_2)}2^{-n(I(X_1,X_2;Y|Z,U,S)  - \delta_6(\epsilon)},
		\end{align}
		since for $j\neq 1, k\neq 1$, the codewords corresponding to $j,k$ are independent of $Y^{n(b)}$ given $S^{n(b)}$ and $U^n(L^{(b-1)})$.\\
		Case \ref{Cases - Packing lemma, DM-MAC - Case 3} by 
		\begin{align}
			\sum_{k>1}\Pr\left[\tilde{E}_{L^{(b-1)},1,k}(b)|\mathrm{E}_5^c{(b+1)},\tilde{E}_1(B)\right] & \leq 2^{nR_2}2^{-n(I(X_2;Y|X_1,U,S) - \delta_7(\epsilon)},
		\end{align}
		because $X_2^n(k|L^{(b-1)})$ is independent of $Y^{n(b)}$ given $(S^{n(b)},U^n(L^{(b-1)}),X_1^n(1|1,K^{(b)},L^{(b-1)},S^{n(b)}))$,
		and case \ref{Cases - Packing lemma, DM-MAC - Case 4} by 
		\begin{align}
			\sum_{j>1}\Pr\left[\tilde{E}_{L^{(b-1)},j,1}(b)|\mathrm{E}_5^c{(b+1)},\tilde{E}_1(B)\right] & \leq 2^{nR_1^{\dprime}}2^{-n(I(X_1;Y|X_2,Z,U,S) - \delta_8(\epsilon)}.
		\end{align}
		since $X_1^n(1|1,K^{(b)},L^{(b-1)})$ is independent of $Y^{n(b)}$ given $(S^{n(b)},U^n(L^{(b-1)}),X_2^n(k|L^{(b-1)}),S^{n(b)})$.
		Thus, the above probabilities goes to zero as $n\to\infty$ if
		\begin{subequations}
			\begin{align}
			R_B+R_1^\dprime+R_2 &< I(X_1,X_2;Y|S) - \delta_5(\epsilon) + I(U;S) - \delta_4(\epsilon) \\
			R_1^\dprime + R_2 &< I(X_1,X_2;Y|U,S,Z) - \delta_6(\epsilon) \\
			R_2 &< I(X_2;Y|U,S,X_1) - \delta_7(\epsilon) \\
			R_1^\dprime &< I(X_1;Y|Z,U,S,X_2) - \delta_8(\epsilon)
		\end{align}
		\end{subequations}
		\item[--] \emph{Event $\left[\mathrm{E}_5^{(b)}|\mathrm{E}_3^c(b),\mathrm{E}_4^c(b),\mathrm{E}_5^c{(b+1)},\tilde{E}_1^c(B)\right]$:} \\
		The probability of this event is zero since $\mathrm{E}_5^{(b)}\cap \mathrm{E}_4^c(b) = \phi $.
	\end{itemize}
	Henceforth, we derived with the following bounds
	\begin{subequations}
		\begin{align}
		\tilde {R}_1 &< H(Z|U,S) - \delta_2(\epsilon)\\
		R_B &> \tilde{R}_1 + \delta_1(\epsilon) \\
		\tilde{R}_1 &> I(U;S) + \Delta_n \\
		R_1^\prime + \tilde{R}_1 &< H(Z|U,S)-\delta_3(\epsilon) \\
		R_1^\prime + \tilde{R}_1 &< R_B \\
		R_B+R_1^\dprime+R_2 &< I(X_1,X_2;Y|S) + I(U;S) - \delta_4(\epsilon) - \delta_5(\epsilon) \\
		R_1^\dprime + R_2 &< I(X_1,X_2;Y|U,S,Z) - \delta_6(\epsilon) \\
		R_2 &< I(X_2;Y|U,S,X_1) - \delta_7(\epsilon) \\
		R_1^\dprime &< I(X_1;Y|Z,U,S,X_2) - \delta_8(\epsilon)
	\end{align}
	\end{subequations}
	together with the identity $R_1=R_1^\prime+R_1^\dprime$ and non-negativity of all rates. Applying Fourier-Motzkin elimination (can be done using \cite{gattegno2016fourier}) to eliminate $R_1^\prime,R_1^\dprime,\tilde{R}_1$ and $R_B$, yields
	\begin{subequations}
		\begin{align}
		I(U;S)&<H(Z|U,S) \\
		R_2&<I(X_2;Y|U,S,X_1) \\
		R_1&<I(X_1;Y|U,S,X_2,Z)+H(Z|U,S)-I(U;S) \\
		R_1+R_2&<I(X_1,X_2;Y|S) \\
		R_1+R_2&<I(X_1,X_2;Y|U,S,Z)+H(Z|U,S)-I(U;S)
	\end{align}
	\end{subequations}
	This closes the proof for the direct part. \hfill$\square$
	\subsection{Converse}\label{Subsection: Converse for MAC one state}
	Assuming that the rate pair $(R_1,R_2)$ is achievable, 
	\begin{align}
	P_e(\mathcal{C}_n) \leq \epsilon.
	\end{align}
	By Fano's inequality, we have
	\begin{align}
	H({M}_1,{M}_2|Y^n,S^n) \leq H_b(P_e(\mathcal{C}_n)) + P_e(\mathcal{C}_n)\log\left(|\mathcal{M}_1\times\mathcal{M}_2| -1\right)
	\end{align}
	where $H_b(\cdot)$ is the binary entropy function.
	Define
	\begin{subequations}
		\begin{align}
		\epsilon_n =& \frac{1}{n}\left(H_b(P_e(\mathcal{C}_n)) + P_e(\mathcal{C}_n)\log\left(|\mathcal{M}_1\times\mathcal{M}_2| -1\right)\right) \\
		\leq& \frac{1}{n}H_b(P_e(\mathcal{C}_n)) + P_e(\mathcal{C}_n)(R_1 + R_2)
		\end{align}
	\end{subequations}
	Note that $\epsilon_n\to0$ when $\epsilon \to 0$. 
	To show that the region in \eqref{Equation: MAC non-causal capacity region} is an outer bound, we first identify the auxiliary random variable $U_i\defeq(Z^{i-1},S^{i-1})$.\\
	Consider
	\begin{subequations}
		\begin{align}
		0 &\leq H(Z^n|S^n)  \\
		&= H(Z^n,S^n) - H(S^n) \\
		&\stackrel{(a)}{=} \sum_{i=1}^{n} H(Z_i,S_i|Z^{i-1},S^{i-1}) - H(S_i)\\ 
		& = \sum_{i=1}^{n} H(Z_i|S_i,Z^{i-1},S^{i-1}) - \left((H(S_i) - H(S_i|S^{i-1},Z^{i-1})\right) \\
		& \stackrel{(b)}{=} \sum_{i=1}^{n} H(Z_i|S_i,U_i) - (H(S_i) - H(S_i|U_i)) \\
		& = \sum_{i=1}^{n} H(Z_i|S_i,U_i) - I(S_i;U_i) \\
		& \stackrel{(c)}{=} n(H(Z_Q|U_Q,S_Q,Q) - I(S_Q;U_Q|Q))
		\end{align}
	\end{subequations}
	where: \\
	(a) - since $S^n$ is i.i.d, \\
	(b) - by definition of $U_i$, \\
	(c) - by setting $Q$ to be a time sharing-random variable, $Q\sim U[1:n]$. \\
	\ \\
	Therefore, we have shown that 
	\begin{subequations}\label{Equation: DM-MAC converse I(US) inequality}
		\begin{align}
		I(U_Q;S_Q|Q) &\leq H(Z_Q|U_Q,S_Q,Q) \\
		H(Z^n|S^n) &= n\left(H(Z_Q|U_Q,S_Q,Q) - I(U_Q;S_Q|Q)\right).
		\end{align}
	\end{subequations}
	An upper bound on $R_1$ is established as follows
	\begin{subequations}
		\begin{align}
		nR_1 &= H(M_1) \\
		&\stackrel{(a)}{=} H(M_1|S^n) \\
		&\stackrel{(b)}{=} H(M_1,Z^n|S^n) \\
		&= H(Z^n|S^n) + H(M_1|Z^n,S^n) \\
		&\stackrel{(c)}{=} H(Z^n|S^n) + H(M_1|Z^n,S^n,M_2) \\
		&\stackrel{}{=} H(Z^n|S^n) + H(M_1|Z^n,S^n,M_2) - H(M_1|Z^n,S^n,M_2,Y^n) + H(M_1|Z^n,S^n,M_2,Y^n)\\
		&\stackrel{(d)}{\leq} H(Z^n|S^n) + I(M_1;Y^n|Z^n,S^n,M_2) + n\epsilon_n \\
		& \stackrel{}{=} H(Z^n|S^n) + \sum_{i=1}^{n}I(M_1;Y_i|Z^n,S^n,Y^{i-1},M_2) + n\epsilon_n \\
		&\stackrel{(e)}{=} H(Z^n|S^n) + \sum_{i=1}^{n}I(M_1;Y_i|Z^n,S^n,Y^{i-1},M_2,X_{2,i}) + n\epsilon_n \\
		&\stackrel{}{\leq} H(Z^n|S^n) + \sum_{i=1}^{n}I(M_1,M_2,X_{1,i},Y^{i-1},Z_{i+1}^n,S_{i+1}^n;Y_i|Z^{i},S^i,X_{2,i}) + n\epsilon_n \\
		&\stackrel{(f)}{=} H(Z^n|S^n) + \sum_{i=1}^{n}I(X_{1,i};Y_i|Z_{i},S_i,X_{2,i},U_i) + n\epsilon_n \\
		&\stackrel{(g)}{=} n\left(H(Z_Q|U_Q,S_Q,Q) - I(U_Q;S_Q|Q)+I(X_{1,Q};Y_Q|X_{2,Q},S_Q,Z_Q,U_Q,Q)  + \epsilon_n\right)
		\end{align}
	\end{subequations}
	where: \\
	(a) - since $M_1 \independent S^n$, \\
	(b) - since $Z^n$ is a function of $M_1$ and $S^n$, \\
	(c) - since $M_2 \independent (M_1,S^n,Z^n)$, \\
	(d) - by Fano's inequality, \\
	(e) - since $X_{2,i}$ is a function of $M_2$ and $Z^{i-1}$, \\
	(f) - due to the Markov chain $(M_1,M_2,Z_{i+1}^n,S_{i+1}^n,Y^{i-1}) \Markov (X_{1,i},X_{2,i},S_i,U_i,Z_i) \Markov Y_i$, \\
	(g) - follows from (\ref{Equation: DM-MAC converse I(US) inequality}) and time-sharing variable $Q$. \\
	Applying similar arguments, we get an upper bound for $R_2$
	\begin{subequations}
		\begin{align}
		nR_2 &= H(M_2) \\
		& = H(M_2|S^n,M_1) \\
		& = H(M_2|S^n,M_1) - H(M_2|S^n,M_1,Y^n) + H(M_2|S^n,M_1,Y^n)\\
		& \leq I(M_2;Y^n|S^n,M_1) + n\epsilon_n\\
		& =\sum_{i=1}^{n} I(M_2;Y_i|S^n,M_1,Y^{i-1}) + n\epsilon_n			\\
		& =\sum_{i=1}^{n} I(M_2,X_{2,i};Y_i|S^n,M_1,Y^{i-1},Z^{i},X_{1,i}) + n\epsilon_n	\\
		& \leq \sum_{i=1}^{n} I(M_2,M_1,Y^{i-1},S_{i+1}^n,X_{2,i};Y_i|S^i,Z^{i-1},X_{1,i}) + n\epsilon_n			\\
		& \stackrel{(h)}{=} \sum_{i=1}^{n} I(X_{2,i};Y_i|S_i,X_{1,i},U_i) + n\epsilon_n			\\
		& = n(I(X_{2,Q};Y_Q|S_Q,X_{1,Q},U_Q,Q) + \epsilon_n)
		\end{align}
	\end{subequations}
	where: \\
	(h) follows since $(M_2,M_1,Y^{i-1},S_{i+1}^n) \Markov(X_{1,i},X_{2,i},S_i,U_i) \Markov Y_i$ is a Markov chain. \\
	The first upper bound for the sum-rate is:
	\begin{subequations}
		\begin{align}
		n\left(R_1+R_2\right)&= H(M_1)+H(M_2) \\
		&= H(M_1,M_2) \\
		&= H(M_1,M_2|S^n) \\
		& = H(Z^n|S^n) + H(M_1,M_2|S^n,Z^n) \\
		&\leq H(Z^n|S^n) + I(M_1,M_2;Y^n|S^n,Z^n) + n\epsilon_n \\
		&= H(Z^n|S^n) + \sum_{i=1}^{n} I(M_1,M_2;Y_i|S^n,Z^n,Y^{i-1}) + n\epsilon_n \\
		&= H(Z^n|S^n) + \sum_{i=1}^{n} I(M_1,M_2,X_{1,i},X_{2,i};Y_i|S^n,Z^n,Y^{i-1}) + n\epsilon_n \\
		&\leq  H(Z^n|S^n) + \sum_{i=1}^{n} I(M_1,M_2,Y^{i-1},S_{i+1}^n,Z_{i+1}^n,X_{1,i},X_{2,i};Y_i|S^i,Z^i) + n\epsilon_n \\
		& = H(Z^n|S^n) + \sum_{i=1}^{n} I(X_{1,i},X_{2,i};Y_i|S_i,Z_i,U_i) + n\epsilon_n \\
		& = n\left(H(Z_Q|U_Q,S_Q,Q) - I(U_Q;S_Q|Q) + I(X_{1,Q},X_{2,Q};Y_Q|S_Q,U_Q,Z_Q,Q) + \epsilon_n\right)
		\end{align}
	\end{subequations}
	and the second upper bound by:
	\begin{subequations}
		\begin{align}
		n\left(R_1+R_2\right)&= H(M_1,M_2) \\
		&= H(M_1,M_2|S^n) \\
		&= H(M_1,M_2,Z^n|S^n) \\
		&\leq I(M_1,M_2;Y^n|S^n) + n\epsilon_n \\
		&= \sum_{i=1}^{n} I(M_1,M_2;Y_i|Y^{i-1},S^n) + n\epsilon_n \\
		&\leq  \sum_{i=1}^{n} I(M_1,M_2,Y^{i-1},S^n_{i+1},X_{1,i},X_{2,i};Y_i|S_i) + n\epsilon_n \\
		& = \sum_{i=1}^{n} I(X_{1,i},X_{2,i};Y_i|S_i) + n\epsilon_n \\
		& = n\left(I(X_{1,Q},X_{2,Q};Y_Q|S_Q,Q) + \epsilon_n\right) \\
		& \leq n\left(I(X_{1,Q},X_{2,Q};Y_Q|S_Q) + \epsilon_n\right)
		\end{align}
	\end{subequations}
	where the last inequality is due to the Markov chain $Q\Markov (X_{1,Q},X_{2,Q},S_Q)\Markov Y_Q$.
	
	We note that the following conditions must hold:
	\begin{itemize}
		\item $S_Q$ is independent of $Q$, and $p_{S_Q}(s)= p_S(s)$.
		\item The Markov $(X_{1,Q},S_Q) \Markov (Q,U_Q) \Markov X_{2,Q}$ holds.
		\item $P_{Y_Q|X_{1,Q},X_{2,Q},S_Q,U_Q,Z_Q,Q}(y|x_1,x_2,s,u,z,q)$ is equal to $p_{Y|X_1,X_2,S}(y|x_1,x_2,s)$.
		\item $Z_Q = z(X_{1,Q},S_Q)$.
	\end{itemize}
	together with the Markov chains
	\begin{subequations}\label{Equation: DM-MAC Markov chains in Converse}
		\begin{align}
		&(M_1,M_2,Z_{i+1}^n,S_{i+1}^n,Y^{i-1}) \Markov (X_{1,i},X_{2,i},S^{i},Z^{i}) \Markov Y_i \\
		& (M_1,M_2,S_{i+1}^n,Y^{i-1}) \Markov (X_{1,i},X_{2,i},S^{i}) \Markov Y_i
		\end{align}		
	\end{subequations}
	
	The first condition holds since $S^n$ is i.i.d. The fourth condition holds since for each $i\in \left[1:n\right]$, $Z_i=z(X_i,S_i)$.
	To prove the second condition, consider
	\begin{subequations}
		\begin{align}
		p(z^{i-1},s^i,x_{1,i},x_{2,i}) &= \sum_{m_1,m_2} p(m_1)p(m_2)p(s^{i-1})p(s_i)p(z^{i-1}|s^i,m_1)p(x_{1,i}|z^{i-1},s^i,m_1)1(x_{2,i}|m_2,z^{i-1}) \\
		& = \sum_{m_1,m_2} p(m_1)p(s^{i-1})p(s_i)p(z^{i-1}|s^i,m_1)p(x_{1,i}|z^{i-1},s^i,m_1)p(m_2)1(x_{2,i}|m_2,z^{i-1}) \\
		& = \sum_{m_1,m_2} p(s^{i-1})p(s_i)p(z^{i-1},x_{1,i},m_1|s^i)p(x_{2,i},m_2|z^{i-1}) \\
		& =  p(s^{i-1})p(s_i)\sum_{(m_1)\in \mathcal{M}_1}p(z^{i-1},x_{1,i},m_1|s^i)\sum_{(m_2)\in \mathcal{M}_2}p(x_{2,i},m_2|z^{i-1}) \\
		& = p(s^{i-1})p(s_i)p(z^{i-1},x_{1,i}|s^i)p(x_{2,i}|z^{i-1}) 
		\end{align}
	\end{subequations}
	which proves that for each $i\in\left[1:n\right]$, the Markov $(X_{1,i},S_i) \Markov (Z^{i-1},S^{i-1}) \Markov X_{2,i}$ holds, and therefore the Markov in the second condition holds. The third condition is due to the memoryless property of the channel and that for random time $Q$, the channel's input are $(X_{1,Q},X_{2,Q},S_Q)$. To see this, consider the PMF of the random variables, that is given by
	\begin{subequations}
		\begin{align}
		p(m_1,m_2,s^n,x_1^n,&z^n,x_2^n,y^n) = \nonumber\\
		&p(m_1)p(m_2)\prod_{i=1}^{n}p(s_i)\prod_{i=1}^{n}1(x_{1,i}|m_1,s^n) 1(z_i|x_{1,i},s_i)1(x_{2,i}|m_2,z^{i-1})p(y_i|x_{1,i},x_{2,i},s_i)
		\end{align}
	\end{subequations}
	It is easy to verify that the Markov chains in Eq. \ref{Equation: DM-MAC Markov chains in Converse} also hold due to this distribution.
	
	Note that $I(U_Q;S_Q|Q) = I(U_Q,Q;S_Q)$ due to the first condition. Let $U=(U_Q,Q), S=S_Q, X_1=X_{1,Q},X_2=X_{2,Q},Z=Z_Q$ and $Y=Y_Q$. Thus, the rate-bounds become
	\begin{subequations}
		\begin{align}
		R_1 &\leq I(X_1;Y|S,Z,U,X_2) + H(Z|U,S) - I(U;S) + \epsilon_n \\
		R_2 &\leq I(X_2;Y|S,U,X_1) + \epsilon_n\\
		R_1+R_2 &\leq I(X_1,X_2;Y|U,Z,S) + H(Z|U,S) - I(U;S) + \epsilon_n\\
		R_1+R_2 & \leq I(X_1,X_2;Y|S)+ \epsilon_n
		\end{align}
	\end{subequations}
	with PMF that factorizes as
	\begin{align}
	p_S(s)p_{U|S}(u|s)p_{X_1|U,S}(x_1|u,s)\mathds{1}\left[z=z(x,s)\right]p_{X_2|U}(x_2|u) p_{Y|X_1,X_2,S}(y|x_1,x_2,s)
	\end{align}
	This completes the proof for the converse part. \hfill $\square$
	\section{Proof for Theorem \ref{Theorem: MAC One Side Cribbing non-causal Capacity, two components}}\label{Section: Proof for MAC with two states}
	\subsection{Direct}\label{Subsection: Achievability proof for MAC with two states}
	\begin{figure*}[t!]
		\begin{subequations}\label{Equation: DM-MAC two states typicality test}
			\begin{align}
			\bigg(s_1^{n(b)},s_2^{n(b)},&u^n(\hat{l}^{(b-1)}),x_1^n(\hat{m}_1^{\dprime(b)}|\hat{m}_1^{\prime(b)},\hat{k}^{(b)},\hat{l}^{(b-1)},s_1^{n(b)}), z^n(\hat{m}_1^{\prime},\hat{k}|\hat{l}^{(b-1)},s_1^{n(b)}),\dots \nonumber\\
			&x_2^n(\hat{m}_2^{(b)}|\hat{l}^{(b-1)},s_2^{n(b)}),y^{n(b)}\bigg)\in \Typical_\epsilon(p_{S_1,S_2}P_{U,X_1|S_1}P_{X_2|U,S_2}P_{Y,Z|X_1,X_2,S_1,S_2})
			\end{align}
		\end{subequations}
		\hrule
	\end{figure*}
	The achievability part of Theorem \ref{Theorem: MAC One Side Cribbing non-causal Capacity, two components} is based on previous section, with additional operation at Encoder 2. To avoid unnecessary repetitions, we only provide the differences in the achievability part relative to that in the previous section.
	\par \emph{Codebook generation:} Draw a cooperative-bin function $\text{bin}(z^n)\sim\text{Unif}[1:2^{nR_B^\prime}]$ for all $z^n\in\mathcal{Z}^n$. Draw $2^{n(R_B^\prime + R_B^\dprime)}$ sequences $u^{n(b)}(l^{\prime},l^{\dprime})$ for $l^{\prime(b-1)}\in[1:2^{nR_B^\prime}]$ and $l^{\dprime}\in[1:2^{nR_B^\dprime}]$, each one is distributed according to $\prod_{i=1}^{n}P_U(u_i(l^\prime,l^\dprime))$. For each $l^\prime,l^\dprime$ and $s_1^{n}$, draw $2^{n(R_1^\prime+\tilde{R})}$ sequences $z^n(m_1^\prime,k|l^{\prime},l^{\dprime},s_1^n)$ distributed according to $\prod_{i=1}^{n} P_{Z|U,S_1}(z_i|u_i(l^{\prime},l^{\dprime}),s_{1,i})$. Then, draw $2^{nR_1^\dprime}$ codewords $x_1^n(m_1^\dprime|m_1^\prime,k,l^{\prime},l^{\dprime},s_1^n)$ distributed according to $\prod_{i=1}^{n}P_{X_1|U,S_1,Z}\left(x_{1,i}|u_i(l^{\prime},l^{\dprime}),s_1^n,z_i(m_1^\prime,k|l^{\prime},l^{\dprime},s_1^n)\right)$ and $2^{nR_2}$ codewords $x_2^n(m_2|l^{\prime},l^{\dprime},s_2^n)$ distributed according to $\prod_{i=1}^{n}P_{X_2|U,S_2}\left(x_{2,i}|u_i(l^{\prime},l^{\dprime}),s_{2,i}\right)$. Set $l^{\prime(0)}=l^{\dprime(0)}=m_1^{\prime(1)}=m_1^{\dprime(1)}=m_2^{(1)}=1$ for beginning the transmission and $k^{(B)}=m_1^{\prime(B)}=m_1^{\dprime(B)}=m_2^{(B)}=1$ for ending the transmission. This setting will result in average rates $\bar{R_1} = \frac{B-2}{B}R_1$ and $\bar{R}_2 = \frac{B-2}{B}R_2$; the average rates can be close to $R_1$ and $R_2$ by taking sufficiently large $B$. 
	Next, we describe the transmission at block $b$. Assume that from previous operation, $l^{\prime(b-1)}$ and $l^{\dprime(b-1)}$ are known at both encoders.
	\begin{figure}[t]
		\centering
		\psfragscanon
		\psfragfig*[mode=nonstop,scale=0.9]{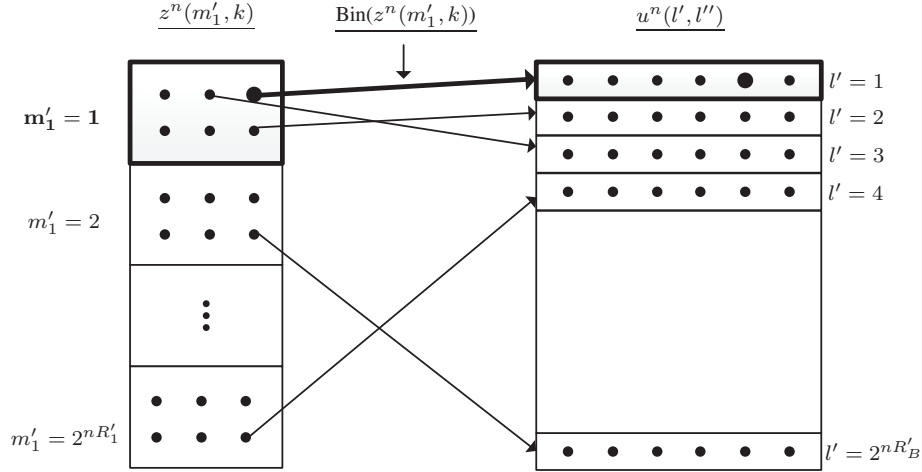}{
			\psfrag{A}[][][1]{\footnotesize $\mathbf{m_1^\prime=1}$}
			\psfrag{B}[][][1]{\footnotesize $m_1^\prime=2$}
			\psfrag{C}[][][1]{\footnotesize $m_1^\prime=2^{nR_1^\prime}$}
			\psfrag{D}[][][1]{\footnotesize $l^\prime = 1$}
			\psfrag{E}[][][1]{\footnotesize $l^\prime = 2$}
			\psfrag{F}[][][1]{\footnotesize $l^\prime = 3$}
			\psfrag{G}[][][1]{\footnotesize $l^\prime = 4$}
			\psfrag{H}[][][1]{\hspace{5mm}\footnotesize $l^\prime = 2^{nR_B^\prime}$}
			\psfrag{I}[][][1]{\footnotesize \underline{$z^n(m_1^{\prime},k)$}}
			\psfrag{J}[][][1]{\footnotesize \underline{Bin($z^n(m_1^{\prime},k)$)}}
			\psfrag{K}[][][1]{\footnotesize \underline{$u^n(l^\prime,l^{\dprime})$}}
		}
		\psfragscanoff
		\caption{Choosing a sequence $z^n$ that points toward a bin, that contains a coordinated sequence $u^n$. The thick dots are the chosen sequences.}
		\label{Figure: Select Bin-Forward with Superbins}
	\end{figure}
	\par \emph{Encoder 1:} Given $m_1^{\prime(b)}$, look for $k^{(b)}$ such that there exist $\tilde{l}^{\dprime(b)}$ that satisfies
	\begin{align}
	\left(u^{n(b+1)}\left(\tilde{l}^{\prime(b)},\tilde{l}^{\dprime(b)}\right),s_1^{n(b+1)}\right)\in\Typical_{\epsilon}(P_{S_1,U}),
	\end{align}
	where $\tilde{l}^{\prime(b)} = \text{bin}\left(z^{n(b)}(m_1^{\prime(b)},k^{(b)}|l^{\prime(b-1)},l^{\dprime(b-1)},s_1^{n(b)})\right)$.
	
	\par This procedure is illustrated in Figure. \ref{Figure: Select Bin-Forward with Superbins}. The cooperative bin index is a superbin, that contains several $u^n$ sequences. The selected superbin contains a sequence $u^n$ that is coordinated with the states. 
	\par \emph{Encoder 2:} At the end of each block $(b-1)$, the superbin index $l^{b-1}$ is known from the cribbed sequence $z^{n(b-1)}$. First, look for the first $\tilde{l}^{\dprime(b-1)}$ s.t. $\left(u^{n(b)}(l^{\prime(b-1)},l^{\dprime(b-1)}),s_2^{n(b)}\right)\in\Typical_\epsilon(P_{S_2,U})$. Then, the encoder sends $x_2^n(m_2^{(b)}|l^{\prime(b-1)},l^{\dprime(b-1)},s_2^n)$.
	
	\par \emph{Decoder:} The decoding is done backwards. Assume that $l^{(b)}=(l^{\prime(b)},l^{\dprime(b)})$ is known from previous decoding operations. 
	\begin{enumerate}
		\item For each $l^{\prime(b-1)}$, find ${\tilde{l}}^{\dprime(b-1)}(l^{\prime(b-1)},s_2^{n(b)})$ the same way that encoder 2 does. Then, find $\hat{m}_1^{\prime(b)}\left(l^{\prime(b-1)},s_2^{n(b)}\right)$ and $\hat{k}^{(b)}\left(l^{\prime(b-1)},s_2^{n(b)}\right)$ s.t. $\text{bin}\left(\hat{m}_1^{\prime(b)},\hat{k}^{(b)}|l^{\prime(b-1)},{\tilde{l}}^{\dprime(b-1)},s_1^{n(b)}\right) = l^{(b)}$. If there are multiple functions that satisfies the above, choose one uniformly. Note that there are total of $2^{nR_B^\prime}$ tuples of functions, since we choose exactly one tuple for each $l^{\prime(b-1)}\in[1:2^{nR_u^\prime}]$. 
		\item Look for $\left(\hat{l}^{\prime(b-1)},\hat{m}_1^{\dprime(b)},\hat{m}_2^{(b)}\right)$ such that \eqref{Equation: DM-MAC two states typicality test} is satisfied. Here we denote $\hat{l}^{(b-1)}=\hat{l}^{\prime(b-1)},\hat{l}^{\dprime(b-1)}$ for abbreviation.
	\end{enumerate}
	Note at at block $B$, the decoder knows the messages and therefore it needs only to find $l^{(B-1)}$ according to the first operation.
	\par \emph{Error analysis:}
	Without loss of generality assume all messages $m_1^{\prime(b)},m_1^{\dprime(b)},m_2^{(b)}$ are equal to $1$ for all $b\in[1:B]$. We begin with the event of encoding error. Recall that according to Lemma \ref{Lemma: indirect covering bins} we can ensure that we will see approximately $R_B^{\dprime} + \tilde{R}$ different indexes by taking $\tilde{R} \leq H(Z|U,S_1)$ and $\tilde{R} < R_B^{\prime}$. Thus, the existence of a sequence $U^{n(b+1)}$ that is coordinated with $S_1^{n(b+1)}$ is also ensured by taking $I(U;S_1)< R_B^{\dprime} + \tilde{R}$. Moreover, it follows from Markov lemma \cite[Lemma 12.1]{el2011network} that $\left(U^{(b+1)},S_2^{n(b+1)}\right)\in\Typical_\epsilon$ with high probability (goes to $1$ when $n$ goes to infinity). Denote the selected superbin of the next block by $L^{\prime(b)}$ and the selected index in the bin by $L^{\dprime(b)}$. 
	At Encoder 2, we ensure that there exist only one $l^{\dprime(b)}$ such that the sequence $U^n(L^{\prime(b)},l^{\dprime(b)})$ is jointly typical with $S_2^{n(b+1)}$; this is done by taking $R_B^{\dprime} < I(U;S_2)$.
	At the decoder, an error occurs if equation \eqref{Equation: DM-MAC two states typicality test} is satisfied by $(\hat{l}^{\prime(b-1)},\hat{m}_1^{\prime(b)},\hat{m}_1^{\dprime(b)},\hat{m}_2^{(b)})\neq (L^{(b-1)},1,1,1)$. This event is bounded by the union of the following events:
	\begin{enumerate}
		\item There exist $(\hat{m}_1^{\prime(b)}, \hat{k}^{(b)})\neq (1,K^{(b)})$ such that $\text{bin}\left(Z^{n(b)}(\hat{m}_1^{\prime(b)}, k^{(b)}|L^{(b-1)},S_1^{n(b)}) \right)= L^{(b)}$.
		\item $\left(\hat{l}^{\prime(b-1)},\hat{m}_1^{\dprime(b)},\hat{m}_2^{(b)}\right)= (\neq L^{(b-1)},*,*)$
		\item $\left(\hat{l}^{\prime(b-1)},\hat{m}_1^{\dprime(b)},\hat{m}_2^{(b)}\right)= (L^{(b-1)},>1,>1)$
		\item $\left(\hat{l}^{\prime(b-1)},\hat{m}_1^{\dprime(b)},\hat{m}_2^{(b)}\right)= (L^{(b-1)},1,>1)$
		\item $\left(\hat{l}^{\prime(b-1)},\hat{m}_1^{\dprime(b)},\hat{m}_2^{(b)}\right)= (L^{(b-1)},>1,1)$
	\end{enumerate}
	Following similar steps as in Section \ref{Section: Proof MAC with one state}, a standard application of the packing lemma results in
	\begin{subequations}\label{Equation: MAC two states rates before FME}
		\begin{align}
		R_1^{\prime} + \tilde{R} &< H(Z|U,S_1) \\
		R_1^{\prime} + \tilde{R} &< R_B^{\prime} \\
		R_B^{\prime}+R_1^{\dprime} + R_2 &< I(X_1,X_2;Y|S_1,S_2) + I(U;S_1) \\
		R_1^{\dprime} +R_2 &< I(X_1,X_2;Y | Z_1,U,S_1,S_2) \\
		R_2 &< I(X_2;Y|X_1,U,S_1,S_2) \\
		R_1^{\dprime} &< I(X_1;Y|X_2,Z_1,U,S_1,S_2)
		\end{align}
	and the encoding constraints are
		\begin{align}
		\tilde{R} + R_B^{\dprime} &> I(U;S_1) \\
		\tilde{R} &< H(Z|S_1,U) \\
		\tilde{R} &< R_B^{\prime} \\
		R_B^{\dprime} &< I(U;S_2),
		\end{align}
	\end{subequations}
	Performing FME on \eqref{Equation: MAC two states rates before FME} yields
	\begin{subequations}
		\begin{align}
		R_1 &< I(X_1;Y|Z,U,X_2,S_1,S_2) + H(Z|U,S_1) - I(U;S_1|S_2) \\
		R_2 &< I(X_2;Y|X_1,U,S_1,S_2) \\
		R_1 + R_2 &< I(X_1,X_2;Y|U,Z,S_1,S_2) + H(Z|U,S_1) - I(U;S_1|S_2) \\
		R_1 + R_2 &< I(X_1,X_2;Y|S_1,S_2) \\
		I(U;S_1|S_2) &< H(Z|U,S_1),
		\end{align}
	\end{subequations}
	for all PMFs that factorize as $P_{X_1,U|S_1}P_{X_2|U,S_2}$ and $Z = z(X_1,S_1)$. 
	Note that $I(U;S_1) - I(U;S_2) = I(U;S_1|S_2)$ since $S_2\Markov S_1 \Markov U$ form a Markov chain.
	 \hfill $\square$
	\subsection{Converse}\label{Subsection: Converse proof for MAC with two states}
	Let $U_i \defeq (Z^{i-1},S_1^{i-1},S_{2,i+1}^{n})$.
	\begin{subequations}
		\begin{align}
		H(Z^n|S_1^n,S_2^n) =& H(Z^n,S_1^n,S_2^n) - H(S_1^n,S_2^n) \\
		\stackrel{(a)}{=}& \sum_{i=1}^n \left[ H(Z_i,S_{1,i},S_{2,i}|Z^{i-1},S_1^{i-1},S_{2,i+1}^n) - H(S_{1,i},S_{2,i})\right] \\
		\stackrel{(b)}{=}& \sum_{i=1}^n \left[ H(Z_i|S_{1,i},S_{2,i},U_i)  + H(S_{1,i},S_{2,i}|U_i)- H(S_{1,i},S_{2,i})\right] \\
		\stackrel{}{=}& \sum_{i=1}^n \left[ H(Z_i|S_{1,i},S_{2,i},U_i) - I(U_i;S_{1,i},S_{2,i})\right] \\
		\stackrel{}{\leq}& \sum_{i=1}^n \left[ H(Z_i|S_{1,i},S_{2,i},U_i) - I(U_i;S_{1,i}|S_{2,i})\right] \\
		\stackrel{(c)}{=}& n\left[H(Z_Q|S_{1,Q},S_{2,Q},U_Q,Q) - I(U_Q;S_{1,Q}|S_{2,Q},Q)\right]
		\end{align}
	\end{subequations}
	where (a) follows since $S_1^n$ and $S_2^n$ are drawn i.i.d in pairs, (b) follows by our definition of $U_i$ and (c) is derived by setting $Q\sim\text{Unif}[1:n]$ to be a time sharing random variable.
	\begin{figure}[h!]
		\centering
		\psfragscanon
		\psfragfig*[mode=nonstop,scale=1.1]{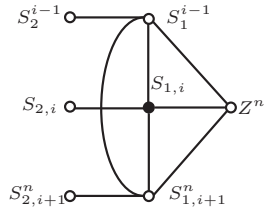}{
			\psfrag{A}[][][1]{\hspace{-2mm} \scriptsize $S_2^{i-1}$}
			\psfrag{B}[][][1]{\hspace{-2mm} \scriptsize $S_{2,i}$}
			\psfrag{C}[][][1]{\hspace{-2mm} \scriptsize $S_{2,i+1}^{n}$}
			\psfrag{D}[][][1]{\hspace{2mm} \scriptsize $S_1^{i-1}$}
			\psfrag{E}[][][1]{\scriptsize $S_{1,i}$}
			\psfrag{F}[][][1]{\hspace{5mm} \scriptsize $S_{1,i+1}^{n}$}
			\psfrag{G}[][][1]{\scriptsize $Z^{n}$}
		}
		\psfragscanoff
		\caption{Proof for Markov chains $S_{2,i}\Markov S_{1,i} \Markov U_i$ and $S_{2,i} \Markov(S_{1,i},U_i)\Markov Z_i$ using an undirected graphical technique \cite{permuter2010twowayhelper}. The undirected graph corresponds the PMF $P(s_1^n,s_2^n,z^n) = P(s_{1}^{i-1},s_{2}^{i-1})P(s_{1,i},s_{2,i})P(s_{1,i+1}^n,s_{2,i+1}^n)P(z^n|s_1^n) $. The Markov chains follows since all paths from $S_{2,i}$ to all other nodes go through $S_{1,i}$.}
		\label{Figure: MAC 2 Components 1 crib - Markovs}
	\end{figure}
	Note that the following Markov chains hold:
	\begin{subequations}\label{Equation: MAC 2 states 1 crib Markov chains}
		\begin{align}
		&S_{2,i}\Markov S_{1,i} \Markov U_i \\
		&S_{2,i} \Markov (S_{1,i},U_i) \Markov Z_i \\
		&(S_{1,i},X_{1,i})\Markov (S_{2,i},U_i) \Markov X_{2,i}
		\end{align}
	\end{subequations}
	Recall that the PMF on $(m_1,m_2,s_1^n,s_2^n,x_{1,i},z^n,x_{2,i})$ is
	\begin{subequations}
		\begin{align}
		P(m_1,m_2,s_1^n,s_2^n,x_{1,i},z^n,x_{2,i}) = & P(m_1)P(m_2)\left[\prod_{i=1}^{n}P(s_{1,i},s_{2,i})\right] 1(x_{1,i},z^n|s_1^n,m_1)1(x_{2,i}|z^{i-1},s_2^n,m_2)
		\end{align}
	\end{subequations}
	Note that $Z^n$ is a deterministic function of $(M_1,S_1^n)$ since $X_1^n$ is. Therefore, the  Markov chain $(S_{1,i},X_{1,i})\Markov (S_{2,i},U_i) \Markov X_{2,i}$ is readily proven from the PMF. As for the other Markovs in \eqref{Equation: MAC 2 states 1 crib Markov chains}, we use an undirected graphical technique in Figure \ref{Figure: MAC 2 Components 1 crib - Markovs}. It is also straightforward to show that $S_{2,Q}\Markov (S_{1,Q},U_{Q},Q) \Markov Z_Q$ holds. Therefore, 
	\begin{align}\label{Equation: DM-MAC with 2 states 1 crib, Zn identity}
	H(Z^n|S_1^n,S_2^n) =& n\left[H(Z_Q|S_{1,Q},S_{2,Q},U_Q,Q) - I(U_Q;S_{1,Q}|S_{2,Q},Q)\right] \\
	=& n\left[H(Z_Q|S_{1,Q},U_Q,Q) - I(U_Q;S_{1,Q}|S_{2,Q},Q)\right] 
	\end{align}
	Note that due to this identity, $I(U_Q;S_{1,Q}|S_{2,Q},Q) \leq H(Z_Q|S_{1,Q},U_{Q},Q)$.
	We proceed to bound $R_1$ and $R_2$. Note that by Fano's inequality,
	\begin{align}
	H(M_1,M_2|Y^n,S_1^n,S_2^n) \leq n\epsilon_n
	\end{align}
	where $\epsilon_n\to 0$ when $n\to\infty$.
	Bounding $R_1$ yields
	\begin{subequations}
		\begin{align}
		nR_1 =& H(M_1) \\
		\stackrel{(a)}{=}& H(M_1|S_1^n,S_2^n) \\
		\stackrel{(b)}{=}& H(M_1,Z^n|S_1^n,S_2^n) \\
		\stackrel{}{=}& H(M_1|Z^n,S_1^n,S_2^n) + H(Z^n|S_1^n,S_2^n) \\
		\stackrel{(c)}{=}& H(M_1|Z^n,S_1^n,S_2^n,M_2) + H(Z^n|S_1^n,S_2^n) \\
		\stackrel{}{\leq}& I(M_1;Y^n|Z^n,S_1^n,S_2^n,M_2) + H(Z^n|S_1^n,S_2^n) +n\epsilon_n\\
		\end{align}
	\end{subequations}
	where: \\
	(a) - follows since $M_1 \independent (S_1^n,S_2^n)$ \\
	(b) - follows since $Z^n=f(M_1,S_1^n)$, \\
	(c) - follows since $M_2 \independent (M_1,Z^n,S_1^n,S_2^n)$.
	It follows that
	\begin{subequations}
		\begin{align}
		I(M_1;Y^n|Z^n,S_1^n,S_2^n,M_2) =& \sum_{i=1}^{n} I(M_1;Y_i|Y^{i-1},Z^n,S_1^n,S_2^n,M_2) \\
		\stackrel{(d)}{=}& \sum_{i=1}^{n} I(M_1,X_{1,i};Y_i|X_{2,i},Y^{i-1},Z^n,S_1^n,S_2^n,M_2) \\
		\stackrel{(e)}{\leq}& \sum_{i=1}^{n} I(X_{1,i};Y_i|X_{2,i},Z^i,S_1^i,S_{2,i}^n) \\
		=& \sum_{i=1}^{n} I(X_{1,i};Y_i|X_{2,i},Z_i,S_{1,i},S_{2,i},U_i) \\
		=& nI(X_{1,Q};Y_Q|X_{2,Q},Z_Q,S_{1,Q},S_{2,Q},U_Q,Q)
		\end{align}
	\end{subequations}
	where (d) follows since $X_{1,i}$ is a function of $(M_1,S_1^n)$ and (e) follows by moving $(M_2,Y^{i-1},Z_{i+1}^n,S_{1,i+1}^n,S_2^{i-1})$ from the conditioning to the left hand side of the mutual information; since the channel is memoryless and without feedback, $(M_2,Y^{i-1},Z_{i+1}^n,S_{1,i+1}^n,S_2^{i-1})\Markov (X_{1,i},X_{2,i},S_{1,i},S_{2,i})\Markov Y_i$ holds.
	
	We derive with the bound
	\begin{align}
	R_1 \leq n\left[I(X_{1,Q};Y_Q|X_{2,Q},Z_Q,S_{1,Q},S_{2,Q},Q) + H(Z_Q|S_{1,Q},U_Q,Q) - I(U_Q;S_{1,Q}|S_{2,Q}) + \epsilon_n\right]
	\end{align}
	
	Following similar steps, we have
	\begin{subequations}
		\begin{align}
		nR_2 =& H(M_2) \\
		=& H(M_2|S_1^n,S_2^n,M_1) \\
		\leq& I(M_2;Y^n|S_1^n,S_2^n,M_1) + n\epsilon_n\\
		=& \sum_{i=1}^{n} I(M_2;Y_i|Y^{i-1},S_1^n,S_2^n,M_1) + n\epsilon_n\\
		=& \sum_{i=1}^{n} I(M_2,X_{2,i};Y_i|Y^{i-1},S_1^n,S_2^n,M_1,X_{1,i}) + n\epsilon_n\\
		\leq& \sum_{i=1}^{n} I(Y^{i-1},S_{1,i+1}^n,S_2^{i-1},M_1,M_2,X_{2,i};Y_i|,S_1^i,S_{2,i}^n,X_{1,i}) + n\epsilon_n\\
		=& \sum_{i=1}^{n} I(X_{2,i};Y_i|S_1^i,S_{2,i}^n,X_{1,i}) + n\epsilon_n\\
		=& nI(X_{2,Q};Y_Q|X_{1,Q},S_{1,Q},S_{2,Q},Q) + n\epsilon_n
		\end{align}
	\end{subequations}
	The sum-rate $R_1+R_2$ is upper bounded by
	\begin{subequations}
		\begin{align}
		n(R_1+R_2) =& H(M_1) + H(M_2) \\
		=& H(M_1,M_2) \\
		=& H(M_1,M_2|S_1^n,S_2^n) \\
		=& H(M_1,M_2,Z^n|S_1^n,S_2^n) \\
		=& H(M_1,M_2|Z^n,S_1^n,S_2^n) +H(Z^n|S_1^n,S_2^n)\\
		\leq& I(M_1,M_2;Y^n|Z^n,S_1^n,S_2^n) + H(Z^n|S_1^n,S_2^n) + n\epsilon_n
		\end{align}
	\end{subequations}
	where
	\begin{subequations}
		\begin{align}
		I(M_1,M_2;Y^n|Z^n,S_1^n,S_2^n) =& \sum_{i=1}^{n}I(M_1,M_2;Y_i|Y^{i-1},Z^n,S_1^n,S_2^n) \\
		\leq& \sum_{i=1}^{n}I(M_1,M_2,S_{1,i+1}^n,S_2^{i-1},Y^{i-1},Z^{i-1},X_{1,i},X_{2,i};Y_i|Z^i,S_1^{i},S_{2,i}^n) \\
		=& \sum_{i=1}^{n}I(X_{1,i},X_{2,i};Y_i|Z_i,S_{1,i},S_{2,i},U_i) \\
		=& nI(X_{1,Q},X_{2,Q};Y_Q|Z_Q,S_{1,Q},S_{2,Q},U_Q,Q)
		\end{align}
	\end{subequations}
	and therefore, it follows from the identity in \eqref{Equation: DM-MAC with 2 states 1 crib, Zn identity} and the above that
	\begin{subequations}
		\begin{align}
		n(R_1+R_2 ) \leq n\big[I(X_{1,Q},X_{2,Q};Y_Q|&Z_Q,S_{1,Q},S_{2,Q},U_Q,Q)  + \nonumber \\
		&H(Z_Q|S_{1,Q},U_Q,Q - I(U_Q;S_{1,Q}|S_{2,Q})) + \epsilon_n \big]
		\end{align}
	\end{subequations}
	and the second upper bound by:
	\begin{subequations}
		\begin{align}
		n\left(R_1+R_2\right)&= H(M_1,M_2) \\
		&= H(M_1,M_2|S_1^n,S_2^n) \\
		&= H(M_1,M_2,Z^n|S_1^n,S_2^n) \\
		&\leq I(M_1,M_2;Y^n|S_1^n,S_2^n) + n\epsilon_n \\
		&= \sum_{i=1}^{n} I(M_1,M_2;Y_i|Y^{i-1},S_1^n,S_2^n) + n\epsilon_n \\
		&\leq  \sum_{i=1}^{n} I(M_1,M_2,Y^{i-1},S_1^{n\setminus i},S_2^{n\setminus i},X_{1,i},X_{2,i};Y_i|S_{1,i},S_{2,i}) + n\epsilon_n \\
		& = \sum_{i=1}^{n} I(X_{1,i},X_{2,i};Y_i|S_{1,i},S_{2,i}) + n\epsilon_n \\
		& = n\left(I(X_{1,Q},X_{2,Q};Y_Q|S_{1,Q},S_{2,Q},Q) + \epsilon_n\right) \\
		& \leq n\left(I(X_{1,Q},X_{2,Q};Y_Q|S_{1,Q},S_{1,Q}) + \epsilon_n\right)
		\end{align}
	\end{subequations}
	where the last inequality is due to the Markov chain $Q\Markov (X_{1,Q},X_{2,Q},S_Q)\Markov Y_Q$.
	Thus, we obtained the following region
	\begin{subequations}\label{Equation: Region of MAC with two states, using Q}
		\begin{align}
		R_1 &<I(X_{1,Q};Y_Q|X_{2,Q},Z_Q,S_{1,Q},S_{2,Q},Q) + H(Z_Q|S_{1,Q},U_Q,Q) - I(U_Q;S_{1,Q}|S_{2,Q}) \\
		R_2 &<I(X_{2,Q};Y_Q|X_{1,Q},S_{1,Q},S_{2,Q},Q) \\
		R_1 + R_2 &<I(X_{1,Q},X_{2,Q};Y_Q|Z_Q,S_{1,Q},S_{2,Q},U_Q,Q)  + H(Z_Q|S_{1,Q},U_Q,Q) - I(U_Q;S_{1,Q}|S_{2,Q})\\
		R_1 + R_2 &<I(X_{1,Q},X_{2,Q};Y_Q|S_{1,Q},S_{1,Q}) \\
		0 &< H(Z_Q|S_{1,Q},U_Q,Q)-I(U_Q;S_{1,Q}|S_{2,Q},Q)
		\end{align}
	\end{subequations}
	for PMFs of the form 
	\begin{align}\label{Equation: PMF for maximization for MAC with two states}
	p(q)p_{S_1,S_2}(s_{1,q},s_{2,q})p(u_q,x_{1,q}|s_{1,q},q)p(x_{2,q}|u_q,s_{2,q})p_{Y|X_1,X_2,S_1,S_2}(y_q|x_{1,q},x_{2,q},s_{1,q},s_{2,q}).
	\end{align}
	Note that the PMF in \eqref{Equation: PMF for maximization for MAC with two states} regarding $S_1,S_2$ and $Y$ follows since the states are i.i.d. and the channel is memoryless and fixed (per state). 
	The rest of the proof (regarding the removal of the time sharing random variable $Q$) is straight-forward using the same steps as in the case of one state component in Appendix \ref{Subsection: Converse for MAC one state}.
	Therefore, by letting $U=(U_Q,Q),X_{1} = X_{1,Q},X_{2} = X_{2,Q},Y=Y_Q,Z=Z_Q,S_1=S_{1,Q}$ and $S_{2,Q}$ we obtain the capacity region in Theorem \ref{Theorem: MAC One Side Cribbing non-causal Capacity, two components}.\hfill $\square$
	\section{Proof for Theorem \ref{Theorem: MAC One Side Cribbing non-causal Capacity, two components, causal}}\label{Section: Proof for MAC with two states, causal}
	The proof for this theorem heavily relies on the proofs from previous sections. The achievability part build on cooperative-bin-forward scheme from section \ref{Section: Proof for MAC with two states} by combining it with instantaneous relaying (a.k.a Shannon strategies). To avoid unnecessary repetition, we only provide the differences on the achievability part and the proofs for Markov chains in the converse.
	
	\par \emph{Achievability:} The codebook generation is done similarly as in \ref{Subsection: Achievability proof for MAC with two states}, with additional conditioning on $Z$ when drawing $x_2^n(m_2^{(b)}|l^{(b-1)},s_2^n)$. Namely, the codebook constructed for Encoder 2 are as follows. For each block $b$, $s_2^n\in\mathcal{S}_2^n$, $z\in \mathcal{Z}$ and $(l^{\prime(b-1)},l^{\dprime(b-1)})$, draw $2^{nR_2}$ codewords \begin{align}
		x_2^n(m_2^{(b)}|z,u^n(l^{\prime(b-1)},l^{\dprime(b-1)}),s_2^n)\sim \prod_{i=1}^{n}p_{X_2|Z,U,S_2}(x_{2,i}|z,u_i(l^{\prime(b-1)},l^{\dprime(b-1)}),s_{2,i})
	\end{align}
	In each transmission block, Encoder 1 performs the same operations as before. Encoder 2 also performs the same operation, but at each time $i$ it transmit $x_{2,i}(m_2^{(b)}|z_i,u^n(l^{\prime(b-1)},l^{\dprime(b-1)}),s_2^n)$. The decoder performs backward decoding as before w.r.t. the new codebook. All other operations are preserved and the same error analysis holds. The derivation result in the same achievable rate region, under the new PMF factorization $p_{U,X_1|S_1}p_{X_2|Z,U,S_2}$.
	\par \emph{Converse:} The only difference in the converse compares to that of the previous section is that we need to show the PMF factorization and prove the new Markov chains. The rate bounds on $R_1$ and $R_2$ are the same and obtained using the exact same arguments.
	Continuing the derivation from this point, we need to show that the following Markov chains hold
	\begin{subequations}\label{Equation: MAC 2 states 1 crib Markov chains, causal}
		\begin{align}
		&S_{2,i}\Markov S_{1,i} \Markov U_i \\
		&S_{2,i} \Markov (S_{1,i},U_i) \Markov Z_i \\
		&(S_{1,i},X_{1,i})\Markov (S_{2,i},U_i,Z_i) \Markov X_{2,i}
		\end{align}
	\end{subequations}
	Note that now the PMF of the random variable is 
	\begin{align}\label{Equation: PMF for MAC with two states and causal cribbing}
	p(m_1,m_2,s_1^n,s_2^n,x_{1,i},z^n,x_{2,i}) = & p(m_1)p(m_2)\left[\prod_{i=1}^{n}p(s_{1,i},s_{2,i})\right] 1(x_{1,i},z^n|s_1^n,m_1)1(x_{2,i}|z^{i},s_2^n,m_2)
	\end{align}
	Now $x_{2,i}$ is also a function of $z_i$ and not only $z^{i-1}$.
	Therefore, the first two Markov-chains hold due to the same arguments in the previous section. As for the last Markov, consider
	\begin{subequations}
		\begin{align}
		p(m_1,m_2,s_1^n,s_2^n,x_{1,i},z^n,x_{2,i}) = & p(m_1)p(m_2)\left[\prod_{i=1}^{n}p(s_{1,i},s_{2,i})\right] 1(x_{1,i},z^n|s_1^n,m_1)1(x_{2,i}|z^{i},s_2^n,m_2) \\
		=& p(s_1^{i-1}) p(s_{1,i},s_{2,i})p(s_{2,i+1}^{n})p(s_2^{i-1}|s_1^{i-1})p(s_{1,i+1}^{n}|s_{2,i+1}^{n}) \nonumber \\
		&\times  p(x_{1,i},z^n,m_1|s_1^n)1(x_{2,i},m_2|z^{i},s_2^n) \\
		=& p(s_1^{i-1}) p(s_{1,i},s_{2,i})p(s_{2,i+1}^{n}) p(x_{1,i},z^n,m_1,s_{1,i+1}^{n}|s_1^i,s_{2,i+1}^{n}) \nonumber \\
		&\times p(x_{2,i},m_2,s_2^{i-1}|z^{i},s_{2,i}^n,s_1^{i-1}) \\
		\end{align}
	\end{subequations}
	Summing for  $(m_1,m_2,,z_{i+1}^n,s_2^{i-1},s_{1,i+1}^n)$ results in
	\begin{align}
	 p(s_1^{i-1})p(s_{1,i},s_{2,i})p(s_{2,i+1}^{n})p(x_{1,i},z^i,|s_1^i,s_{2,i+1}^{n})p(x_{2,i}|z^{i},s_{2,i}^n,s_1^{i-1}) 
	\end{align}
	in which $(S_{1,i},X_{1,i})\Markov (S_{2,i},S_1^{i-1},Z^{i-1},S_{2,i+1}^n,Z_i) \Markov X_{2,i}$ is Markov. All other arguments regarding the memoryless property of the channel and the time-sharing random variable $Q$ hold. This concludes the proof of Theorem \ref{Theorem: MAC One Side Cribbing non-causal Capacity, two components, causal}. \hfill $\square$
	\section{Conclusions and final remarks}\label{Section: Conclusion}
	\par Using a variation of the cooperative-bin-forward scheme, we have found the capacity of the SD-RC and MAC with partial cribbing, when non-causal CSI is given to the decoder and one of the transmitters. Remarkably in the both setups only one auxiliary random variable is used for obtaining the capacity region. The same cooperation codeword is designated to play both the roles of common message and compression of the state sequence. It is evident that in the special case of the MAC the non-causal access to the state endowed states compression and, consequently, increased the capacity region.
	\par Cooperative-bin-forward heavily relies on the fact that the link for the cooperation, i.e., the link from the encoder to the relay (or the cribbed signal in the MAC) is deterministic. Since the transmitter can predict and dictate the observed output (by the relay) it can coordinate with the relay based on the same bin index. However, it is not known how the cooperative-bin-forward scheme can be generalized to cases where the link between the encoder and the relay is a general noisy link.
	\appendices
	\section{Proof for indirect covering lemma}\label{Appendix: Covering lemma}
	\par In section \ref{Section: SD-RC capacity} we presented an \emph{indirect} covering lemma. Although we do not actually perform covering in a traditional manner, we do ask for the number of seen bin indexes.
	Namely, we want to bound the following probability
	\begin{align}
		P_e^{(n)} \defeq \Pr\left[\lvert \left\{l:\;\exists k \text{ s.t. Bin}\left(Z^n(k)\right)=l \right\} \rvert < 2^{n(R-\delta_n)}\right] \leq \Delta_n 
	\end{align}
	and ensure that both $\delta_n$ and $\Delta_n$ goes to zero as $n$ goes to infinity.
	\par Assume $v^n\in\Typical_\epsilon(p_V)$ and recall that according to the random experiment, we have
	\begin{align}
		\Pr\left[\{Z^n(k)\}=\{z^n(k)\}_k,\{\text{Bin}(Z^n(k))\}=\{\text{bin}(z^n(k))\}_k|V^n=v^n\right] = \prod_{k=1}^{2^{nR}}p^n_{Z|V}(z^n(k)|v^n)2^{-nR_B}
	\end{align}
	where $p_{Z|V}^n(z^n(k)|v^n) = \prod_{i=1}^{n}p_{Z|V}(z_i(k)|v_i)$. 
	
	\par Define the sets
	\begin{subequations}
		\begin{align}
			\mathcal{D}_1 &\defeq \left\{k:\; \left(Z^n(k),v^n\right)\in\Typical_{\epsilon^\prime}\right\} \\
			\mathcal{D}_2 &\defeq \left\{k:\; Z^n(k)\neq Z^n(j)\quad\forall j\neq k \text{ and } k,j\in\mathcal{D}_1\right\} \\
			\mathcal{D}_3 &\defeq \left\{k:\; \text{Bin}\left(Z^n(k)\right)\neq \text{Bin}\left(Z^n(j)\right)\quad\forall j\neq k \text{ and } k,j\in\mathcal{D}_2\right\}
		\end{align}
	\end{subequations}
	and the events
	\begin{subequations}
		\begin{align}
		\mathrm{E}_1 & \defeq \lvert \mathcal{D}_1 \rvert < 2^{n(R-\delta_n^{(1)})} \\
		\mathrm{E}_2 & \defeq \lvert \mathcal{D}_2 \rvert < 2^{n(R-\delta_n^{(2)})} \\
		\mathrm{E}_3 & \defeq \lvert \mathcal{D}_3 \rvert < 2^{n(R-\delta_n^{(3)})} \\
		\end{align}
	\end{subequations}
	By definition of $E_3$ and law of total probability, it follows that
	\begin{subequations}
		\begin{align}
			P_e^{(n)} \leq& \Pr\left[\mathrm{E}_3|V^n=v^n\right] \\
			\leq& \Pr\left[\mathrm{E}_1|V^n=v^n\right] + \Pr\left[\mathrm{E}_2|\mathrm{E}_1^c,V^n=v^n\right] + \Pr\left[\mathrm{E}_3|\mathrm{E}_2^c,\mathrm{E}_1^c,V^n=v^n\right]
		\end{align}
	\end{subequations}
	We will bound each probability separately.
	\begin{enumerate}
		\item Define $\theta_k = \indicate\left[\left(Z^n(k),v^n\right)\in\Typical_{\epsilon^\prime}(p_{Z,V}) \right]$, and note that
		$\theta_k\stackrel{\text{i.i.d}}{\sim}\text{Bernoully}\left(\rho_n\right)$, where $1-\tilde{\delta}_n \leq \rho_n \leq 1$ and $\tilde{\delta}_n\to0$ as $n\to\infty$.
		Therefore, for any $\delta^\prime > 0$ we have
		\begin{subequations}
		\begin{align}
			\Pr\left[\mathrm{E}_1|V^n=v^n\right] =& \Pr\left[\lvert \mathcal{D}_1 \rvert < 2^{n(R-\delta_n^{(1)})} \vert V^n=v^n\right] \\
			\stackrel{(a)}{=}& \Pr\left[\lvert \mathcal{D}_1 \rvert < 2^{nR}\rho_n(1-\delta^\prime) \vert V^n=v^n\right] \\
			\stackrel{(b)}{\leq}& 2^{-2^{nR}\rho_n \delta^{\prime 2}/2} \\
			= & \Delta_n^{(1)}
		\end{align}
		\end{subequations}
		where: \\
		(a) - by setting $\delta_n^{(1)} \defeq -\frac{1}{n}\log_2\left(\rho_n(1-\delta^\prime)\right) \xrightarrow[n\to\infty]{}0$, \\
		(b) - by Chernoff's inequality \cite[Appendix B]{el2011network}, \\
		and $\Delta_n^{(1)}\to 0$ as $n\to\infty$.
		
		\item We will now deal with $\mathrm{E}_2$. First, note that given $\mathrm{E}_1^c$, we have with probability one that $\lvert \mathbb{D}_1 \rvert > 2^{n(R-\delta_n^{(1)})}$.
		We are interested in $|\mathcal{D}_2|$, so let us define the \emph{normalized} amount of \" bad \" sequences in $\mathcal{D}_1$,
		\begin{align}
			C_2 = \frac{1}{|\mathcal{D}_1|} \sum_{k\in\mathcal{D}_1} \indicate\left[ \exists j \neq k : Z^n(j) = Z^n(k),\;j\in\mathcal{D}_1 \right]
		\end{align}
		By this definition, it follows that $|\mathcal{D}_2| = |\mathcal{D}_1|(1-C_2)$. First, we bound the expected value of $C_2$ by
		\begin{subequations}
			\begin{align}
				\expectation\left[C_2 | V^n=v^n, \mathrm{E}_1^c\right] = & \sum_{d_1}\Pr\left[\mathcal{D}_1 = d_1 | V^n = v^n, \mathrm{E}_1^c\right]	\expectation\left[C_2 | V^n=v^n, \mathrm{E}_1^c,\mathcal{D}_1=d_1\right] \\
				= &\hspace*{-8mm} \sum_{\substack{d_1: |d_1| > 2^{n(R-\delta_n^{(1)})}}}\hspace*{-9mm}\Pr\left[\mathcal{D}_1 = d_1 | V^n = v^n, \mathrm{E}_1^c\right]\frac{1}{|d_1|}\times \nonumber \\
				&\sum_{k\in d_1}\Pr\left[\exists j \neq k : Z^n(j) = Z^n(k),\;j\in\mathcal{D}_1 \vert V^n = v^n, \mathrm{E}_1^c\right] \\
				\stackrel{(c)}{\leq} & \hspace*{-8mm}\sum_{\substack{d_1:  |d_1| > 2^{n(R-\delta_n^{(1)})}}}\hspace*{-9mm}\Pr\left[\mathcal{D}_1 = d_1 | V^n = v^n, \mathrm{E}_1^c\right]\frac{1}{|d_1|}\sum_{k\in d_1} |d_1|2^{-n(H(Z|V)-\epsilon^\prime)} \\
				= & \hspace*{-8mm}\sum_{\substack{d_1:  |d_1| > 2^{n(R-\delta_n^{(1)})}}}\hspace*{-9mm}\Pr\left[\mathcal{D}_1 = d_1 | V^n = v^n, \mathrm{E}_1^c\right]|d_1| 2^{-n(H(Z|V)-\epsilon^\prime + \delta_n^{(1)})} \\
				\leq & 2^{n(R-H(Z|V)+\epsilon^\prime + \delta_n^{(1)})}
			\end{align}
		\end{subequations}
		Therefore, for any $\gamma_1^\prime > 0$ it follows by Markov's inequality that
		\begin{subequations}
			\begin{align}
				\Pr\left[C_2 > 2^{-n\gamma_1^\prime} | E_1^c,V^n=v^n\right] \leq & 2^{n(R - H(Z|V) + \delta_n^{(1)} + \epsilon^\prime +\gamma_1^\prime)} \\
				= & \Delta_n^{(2)}
			\end{align}
		\end{subequations}
		where $\Delta_n^{(2)}\to 0 $ as $n\to \infty$ if $R < H(Z|V)-\gamma_1$ and $\gamma_1 = \delta_n^{(1)} + \epsilon^\prime + \gamma_1^\prime$.
		By setting $\delta_n^{(2)} = \delta_n^{(1)}-\frac{1}{n}\log_2\left(1-2^{-n\gamma_1^\prime}\right)$ we have
		\begin{align}
			\Pr\left[E_2|E_1^c,V^n=v^n\right] \leq \Delta_2^{(n)}.
		\end{align}
		and $\delta_n^{(2)}\to 0$ as $n\to\infty$.
		\item We will follow similar arguments as the previous bound. Define
		\begin{align}
			C_3 = \frac{1}{|\mathcal{D}_2|}\sum_{k\in\mathcal{D}_2} \indicate\left[\exists j \neq k : \;\text{Bin}(Z^n(j))=\text{Bin}(Z^n(k)),\;j\in\mathcal{D}_2\right]
		\end{align}
		and recall that the probability of each bin index is independent of the realization of $\{Z^n(k)\}_k$. It follows that
		\begin{align}
			\expectation\left[C_3|\mathrm{E}_2^c,\mathrm{E}_1^c,V^n=v^n\right] \leq 2^{n(R-R_B + \delta_n^{(2)})}.
		\end{align}
		By Markov's inequality, for any $\gamma_2^\prime > 0$
		\begin{subequations}
			\begin{align}
			\Pr\left[\mathrm{E}_3 | \mathrm{E}_2^c,\mathrm{E}_1^c,V^n=v^n \right] =& \Pr\left[|\mathcal{D}_3| < 2^{n(R-\delta_n^{(3)})} | \mathrm{E}_2^c,\mathrm{E}_1^c,V^n=v^n \right] \\
			=& \Pr\left[C_3 > 2^{-n\gamma_2^\prime} | \mathrm{E}_2^c,\mathrm{E}_1^c,V^n=v^n \right] \\
			\leq & 2^{n(R-R_B+\delta_n^{(2)} + \gamma_2^\prime )} \\
			=& \Delta_n^{(3)}
		\end{align}
		\end{subequations}
		where $\Delta_n^{(3)}\to0$ and $\delta_n^{(3)} =\delta_n^{(2)} -\frac{1}{n}(1-2^{-n\gamma_2^\prime})\to0$ as $n\to\infty$, if $R<H(Z|V) - \gamma_2$ where $\gamma_2 = \delta_n^{(2)} + \gamma_2^\prime$.
	\end{enumerate}
	Finally, for any $\gamma_1,\gamma_2 > $ and $n$ sufficiently large, if
	\begin{subequations}
		\begin{align}
		R < H(Z|V) - \gamma_1 \\
		R < H(Z|V) - \gamma_2
		\end{align}
	\end{subequations}
	then 
	\begin{align}
		P_e^{(n)} \leq \Delta_n^{(1)}+\Delta_n^{(2)}+\Delta_n^{(3)}
	\end{align}
	where $\delta_n^{(i)},\Delta_n^{(i)}$ tends to $0$ when $n\to\infty$ for $i=1,2,3$.\hfill $\square$

	\section{Proofs for special cases of MAC}\label{Appendix: proofs for special cases of MAC}
	The special cases in section \ref{Section: Special cases} are captured by Theorem \ref{Theorem: MAC One Side Cribbing non-causal Capacity, two components}. We restate here the region as a reference for the following derivations. 
	To simplify the derivations, we consider the region for only one state component $S$ which is available only at Encoder 1.
	The capacity region for discrete memoryless MAC with non-causal CSI in Fig. \ref{Figure: DM-MAC with one side cribbing and non-causal states, two states} is given by the set of rate pairs $(R_1,R_2)$ that satisfy
	\begin{subequations}\label{Equation: MAC non-causal capacity region in Appendix}
		\begin{align}
		R_1 &\leq I(X_1;Y|X_2,Z,S,U) + H(Z|S,U) - I(U;S) \\
		R_2 &\leq I(X_2;Y|X_1,S,U) \\
		R_1+R_2 &\leq I(X_1,X_2;Y|Z,S,U) + H(Z|S,U) - I(U;S) \\
		R_1+R_2 & \leq I(X_1,X_2;Y|S) 
		\end{align}
		for PMFs of the form $p_{U|S}p_{X_1|S,U}p_{X_2|U}$, with $Z=z(X_1,S)$, that satisfies
		\begin{align}\label{Equation: Special cases appendix compression inequality}
		I(U;S) &\leq H(Z|U,S),
		\end{align}
	\end{subequations}
	\par \emph{Case A: Multiple Access Channel with states (without cribbing):}
	This case is captured by Theorem \ref{Theorem: MAC One Side Cribbing non-causal Capacity, two components} by setting $z(x_1,s)=0, \;\forall x_1\in\mathcal{X}_1, s\in\mathcal{S}$, since in this configuration there is no cribbing between the encoders. The inequality in \eqref{Equation: Special cases appendix compression inequality} results in $I(U;S)\leq 0$, which enforces $U$ to be independent of $S$. Thus, region in \eqref{Equation: MAC non-causal capacity region in Appendix} becomes 
	\begin{subequations}\label{Equation: MAC with STATES speical case, region 2}
		\begin{align}
		R_1 &\leq I(X_1;Y|S,U,X_2) \\
		R_2 &\leq I(X_2;Y|S,U,X_1) \\
		R_1+R_2 &\leq I(X_1,X_2;Y|S,U) \\
		R_1+R_2 & \leq I(X_1,X_2;Y|S),
		\end{align}
	\end{subequations}
	with PMF of the form $p_Up_{X_1|U,S}p_{X_2|U}$.
	Note that $U\Markov (X_1,X_2,S) \Markov Y$ forms a Markov chain. Therefore, the last inequality is redundant. It also implies that the capacity region in  \eqref{Equation: MAC with STATES speical case, region 2} is outer bounded by \eqref{Equation: MAC with STATES speical case, region 1}; degenerating $U$ achieves that outer bound.
	\par \emph{Case B: State dependent MAC with cooperation:}
	We investigate capacity region for the case of orthogonal cooperation link and channel transmission, as depicted in Fig. \ref{Special cases: Figure - MAC w Conference}. The cooperation link here is \emph{strictly causal} due to the cribbing, i.e., $X_{2,i}=f(M_2,X_{1,p}^n)$.
	First, note that the region in \eqref{Equation: MAC with cooperation, region 1} is an \emph{outer} bound, since it is the capacity region of non-causal cooperation, i.e., when $X_{2,i}=f(M_2,X_{1,p}^n)$. 
	The strictly causal configuration is captured by the cribbing setup when setting $X_1=(X_{1c},X_{1p})$, $Z=X_{1,p}$ and the channel transition PMF to $p_{Y|X_{1c},X_2,S}$.
	Then, the region in \eqref{Equation: MAC non-causal capacity region in Appendix} becomes
	\begin{subequations}\label{Equation: MAC with cooperation, region 2}
		\begin{align}
		R_1 &\leq I(X_{1c};Y|S,X_{1p},U,X_2) + H(X_{1p}|U,S) - I(U;S) \\
		R_2 &\leq I(X_2;Y|S,U,X_{1c},X_{1p}) \\
		R_1+R_2 &\leq I(X_{1c},X_2;Y|S,U,X_{1p}) + H(X_{1p}|U,S) - I(U;S) \\
		R_1+R_2 & \leq I(X_{1c},X_{1p},X_2;Y|S) \\
		I(U;S) &\leq H(X_{1c}|U,S).
		\end{align}
	\end{subequations}
	for PMFs of the form $p_{U|S}p_{X_1|U,S}p_{X_2|U}$,
	Note that $I(X_{1c},X_{1p},X_2;Y|S)=I(X_{1c},X_2;Y|S)$ because $X_{1p} \Markov (X_{1c},X_2,S) \Markov Y$ is a Markov chain.
	We identify the rate $H(X_{1p}|U,S)$ as the cooperation rate $R_{12}$.
	Let $p_{X_{1}|U,S} = p_{X_{1p}|U,S}p_{X_{1c}|U,S}$, and $P_{X_{1p}|U=u,S=s}$ be a uniform distribution for every $(u,s)\in\mathcal{U}\times\mathcal{S}$. By doing so, $H(X_{1p}|U,S) = \log_2|X_{1p}|$ and $I(X_1;Y|X_2,U,S,X_{1p}) = I(X_{1c};Y|X_2,U,S)$. The latter holds since $X_{1p}\Markov(X_{1c},X_2,S) \Markov Y$ is a Markov chain and $X_{1c}$ is independent of $X_{1p}$. By denoting $R_{12} = \log_2|\mathcal{X}_{1p}|$, 	the regions in \eqref{Equation: MAC with cooperation, region 1} and \eqref{Equation: MAC with cooperation, region 2} coincide. 
	\par \emph{Case C: Point-to-point with non-causal CSI:}
	First, note that the channel depends only on $X_1$ and $S$. Encoder $1$ sends a message over the channel, and the states are revealed to it non-causally at the beginning of the transmission. Encoder $2$, however, has no message to send; in fact, it cannot send anything over the channel since the channel's output is not affected by $X_2$ at all. Therefore, the rate $R_2$ is $0$.
	This configuration is captured by the MAC when
	\begin{subequations}\label{Special cases: PTP w CSI: equation 1}
		\begin{align}
		R_2&=0 \\
		p_{Y|X_1,X_2,S}&=p_{Y|X_1,S}.
		\end{align}
	\end{subequations}
	Inserting \eqref{Special cases: PTP w CSI: equation 1} into Theorem \ref{Theorem: MAC One Side Cribbing non-causal Capacity, two components} derives with
	\begin{subequations}\label{Special cases: PTP w CSI: equation 2}
		\begin{align}
		R_1 &\leq I(X_1;Y|S,U,Z,X_2) + H(Z|U,S) - I(U;S) \\
		R_1 &\leq I(X_1,X_2;Y|S,U,Z) + H(Z|U,S) - I(U;S) \\
		R_1 & \leq I(X_1,X_2;Y|S) \\
		I(U;S) &\leq H(Z|U,S)
		\end{align}
	\end{subequations}
	with $p_{S,U,X_1}1_{Z|X_1,S}p_{X_2|U}p_{Y|X_1,S}$.
	Due to the Markov chains $X_2\Markov(X_1,S)\Markov Y$ , $X_2\Markov (X_1,U,S,Z) \Markov Y$ and $X_2\Markov (U,S,Z) \Markov Y$, the following identities hold
	\begin{subequations}
		\begin{align}
		I(X_1,X_2;Y|S) =& I(X_1;Y|S) \\
		I(X_1,X_2;Y|S,U,Z) =& I(X_1;Y|S,U,Z,X_2) \\
		I(X_1,X_2;Y|S,U,Z)=&I(X_1;Y|U,Z,S)
		\end{align}
	\end{subequations}
	Therefore, the region in \eqref{Special cases: PTP w CSI: equation 2} reduces to
	\begin{subequations}
		\begin{align}\label{Special cases: PTP w CSI: equation 3}
		R_1 &\leq I(X_1;Y|S,U,Z) + H(Z|U,S) - I(U;S) \\
		R_1 & \leq I(X_1;Y|S) \\
		I(U;S) &\leq H(Z|U,S)
		\end{align}
	\end{subequations}	
	This region is smaller or equal to \eqref{Equation: Case B capacity}; if we drop the first and last inequalities, we get the expression for capacity\footnote{The expressions for the capacity after dropping the constraints are not exactly the same, since the PMF domains are different. However, the capacity coincide, due to the objective and maximization.}. On the other hand, to show that the capacity in \eqref{Equation: Case B capacity} is achievable, degenerate $U$ in \eqref{Special cases: PTP w CSI: equation 3}. The result is that $I(U;S) = 0$, and the last inequality is redundant. Moreover, 
	\begin{subequations}
		\begin{align}
		I(X_1;Y|S,Z) + H(Z|S) =& I(X_1,Z;Y|S) - I(Z;Y|S) + H(Z|S)  \\
		=&I(X_1,Z;Y|S) + H(Z|Y,S), 
		\end{align}
	\end{subequations}
	so  $Z=f(X_1,S)$, thus $I(X_1;Y|S)=I(X_1,Z;Y|S)$. Therefore, the first inequality becomes $R_1 \leq I(X_1;Y|S) + H(Z|S,Y)$, which is also redundant due to the second. 
	\par \emph{Point-to-point with state encoder and output causality constraint:}
	This configuration is captured by the MAC with cribbing, by setting
	\begin{subequations}\label{Equation: Case A - PTP coded substitute}
		\begin{align}
		R_1&=0\\
		p_{Y|X_1,X_2,S}&=p_{Y|X_2,S} \\
		z(x_1,s)&=x_1.
		\end{align}
	\end{subequations}
	The region in \eqref{Equation: MAC non-causal capacity region in Appendix} reduces to
	\begin{subequations}
		\begin{align}
		R_2 &\leq I(X_2;Y|S,U,X_1) \\
		R_2 &\leq I(X_2;Y|S,U,X_1) + H(X_1|U,S) - I(U;S) \\
		R_2 & \leq I(X_1,X_2;Y|S)	\\
		I(U;S) &\leq H(X_1|U,S).
		\end{align}
	\end{subequations}
	with $p_{U,X_1|S}p_{X_2|U}p_{Y|X_2,S}$.
	Notice that $I(X_2;Y|S,U,X_1) \leq I(X_1,X_2,U;Y|S)$, and both $(U,X_1)\Markov(X_2,S)\Markov\nobreak Y$ and $X_1\Markov (X_2,S) \Markov Y$ are Markov chains. Therefore, the third inequality is redundant. Moreover, from the constraint $I(U;S) \leq H(X_1|U,S)$, it follows that $I(X_2;Y|S,U,X_1) \leq I(X_2;Y|U,X_1,S) + H(X_1|U,S) - I(U;S)$; thus, the second inequality is also redundant.
	The Markov chains $Y \Markov (U,S) \Markov X_1$ and $Y \Markov (X_2,S,U) \Markov X_1$ imply that  $I(X_2;Y|S,U,X_1) = I(X_2;Y|S,U)$. 
	Therefore, the region is further reduced to
	\begin{subequations}
		\begin{align}
		R_2 &\leq I(X_2;Y|S,U) \\
		I(U;S) &\leq H(X_1|U,S).
		\end{align}
	\end{subequations}
	Note that $H(X_1|U,S)\leq \log_2|\mathcal{X}_1|$, and therefore, this region is upper bounded by the capacity.
	By taking $P_{X_1|U,S}$ to be uniform distribution for every $(u,s)\in\mathcal{U}\times\mathcal{S}$, the conditional entropy $H(X_1|U,S)$ equals to $\log_2|\mathcal{X}_1|$ and we achieve the capacity.
	\bibliographystyle{IEEEtran}
	\bibliography{ref}
\end{document}